\documentclass[prb,superscriptaddress,amsfonts,amssymb,amsmath,twocolumn,aps,footinbib,shownopacs]{revtex4-2}

\usepackage[pdftex]{graphicx}
\usepackage{subcaption}
\usepackage{ragged2e} 

\usepackage{dsfont}
\usepackage{dcolumn}
\usepackage{bm}
\usepackage{color}
\usepackage{physics}
\usepackage{multirow}
\usepackage{cancel}
\usepackage[pdftex,colorlinks=true, linkcolor=blue,citecolor=blue,filecolor=blue]{hyperref}
\usepackage{verbatim}

\usepackage{mathrsfs}  

\begin{document}

\title{Sambe Approach to Floquet-Lindblad Open Quantum Systems}

\author{Andriani Keliri}
\affiliation{JEIP, UAR 3573 CNRS, Coll\`ege de France, PSL Research University, 11 Place Marcelin Berthelot, 75321 Paris Cedex 05, France}
\author{Marco Schir\`o}
\affiliation{JEIP, UAR 3573 CNRS, Coll\`ege de France, PSL Research University, 11 Place Marcelin Berthelot, 75321 Paris Cedex 05, France}

\date{\today}

\begin{abstract}
We study driven and open quantum systems described by a time-periodic Lindblad master equation. In closed systems, the stroboscopic dynamics can always be described by an effective time-independent Floquet Hamiltonian; this idea is the basis of Floquet engineering. However, in the presence of dissipation, the existence of an effective time-independent Floquet Lindbladian is not guaranteed due to the non-unitary nature of the evolution. Using Floquet theory, we construct a well-defined time-independent Floquet Lindbladian in an extended Sambe-Liouville space, transforming the initial time-dependent problem to a static and non-Hermitian eigenvalue problem. For harmonic driving, we introduce a matrix continued fraction method to nonperturbatively resum multiphoton processes and construct an effective Floquet Lindbladian acting only on the physical Liouville space. Compared to other high-frequency expansions, this method has the advantage of providing the whole infinite series expansion at once. Using a resolvent formalism, we show how to obtain a spectral Floquet representation of correlation functions of an open quantum system. As an application, we consider a dissipating two-level system in a linearly polarized field and calculate its resonance fluorescence spectrum. Furthermore, we consider a parametrically driven quantum dot with pump and loss for which we calculate its spectral function and current-voltage characteristics.
\end{abstract}
\maketitle 	

\section{Introduction} 
The main idea behind Floquet engineering is that for a Hamiltonian which is periodic in time with period $T,$ $H(t+T)=H(t),$ the stroboscopic dynamics is described by an effective, time-independent Hamiltonian, $H_F,$ called the Floquet Hamiltonian. The Floquet Hamiltonian can be defined through the logarithm of the one-cycle evolution operator $U(T)=e^{-iH_F T}.$ One can then try to think of a drive protocol that allows engineering the desired $H_F$, which can have very different properties from the initial $H(t)$~\cite{floquetengineering,Holthaus,highfrequency,basov,Dalibard}. Examples range from ultracold atoms in optical lattices~\cite{eckardt2017colloquium} to solid-state settings, including the opening of bandgaps in light-irradiated graphene~\cite{photovoltaic,graphene1}, which has been recently experimentally observed~\cite{wang2026observation}, inducing topology in topologically-trivial materials~\cite{lindner2011floquet,graphene2,rudner} and realizing discrete-time crystals~\cite{timecrystals,timecrystals1}. In quantum information, protocols for dynamical stabilization and protection of qubits by noise also constitute examples of Floquet engineering~\cite{sweetspots,floquetqubit,sameti2019floquet,nguyen2024programmable,chessari2025unifying}.

At the same time, it is interesting to extend the concept of Floquet engineering to open quantum systems. Firstly, a more realistic description of currently explored quantum platforms has to include the presence of dissipation. Secondly, periodic driving in closed systems generically causes thermalization~\cite{dalessio2014longtime,lazarides,abanin2015exponentially,mori2016rigorous}, while in an open system a non-trivial non-equilibrium steady state might be reached due to the balance between the driving-induced heating and the dissipation-provided cooling~\cite{floquetinopen,TSUJI2024967}. The effect of dissipation on Floquet engineering is therefore relevant for a broad range of applications, from few- to many-body physics~\cite{Seetharam2015controlled,cohen2023reminiscence,riva2026firstprinciplesstudydispersivereadout,ritter2025autonomous}.

A particularly fruitful approach is to consider that after tracing out the degrees of freedom of the environment, the quantum system can be described by a Markovian master equation, where the generator of the non-unitary dynamics is of Lindblad form~\cite{lindblad1976generators,gorini1976completely}. The Lindblad form ensures the  evolution is given by a completely positive and trace-preserving (CPTP) map, so that the obtained density matrix stays physical at all times. It is then natural to ask whether a Floquet Lindbladian can be defined in analogy to closed systems. However, it turns out that the existence of a Floquet Lindbladian is not trivial due to the non-unitary nature of the evolution~\cite{haddadfarshi2015completely,schnell_is_2020}. 

Several approaches have been developed to address this problem. An elegant approach based on the Lie algebraic method proposed by Wei and Norman~\cite{wei_lie_1963} allows writing the exact solution in cases where the time-dependent Lindbladian forms a closed algebra~\cite{chruscinski_general_2010,scopa_exact_2019,bakker_liealgebraic_2020,PhysRevB.111.165131,qvarfort_solving_2025}. However, this method only applies to specific classes of models and can quickly become unmanageable if the underlying Lie algebra is too complex. A more widely used idea, inherited from the closed system literature, is to develop high-frequency expansions, such as the Magnus and van-Vleck expansions~\cite{schnell_high-frequency_2021,mizuta_breakdown_2021,ikeda_nonequilibrium_2021}. These provide systematic perturbative constructions of the Floquet Lindbladian when the driving frequency is the dominant energy scale. While these expansions offer analytical insight, they typically break down at moderate driving frequencies, and the derivation of higher-order terms is of increasing difficulty. Moreover, whether the expansion is a valid Lindbladian (in the sense that it generates CPTP dynamics) depends on the choice of method and reference frame~\cite{schnell_high-frequency_2021} and one has to check the regions of validity for each specific model.

A common approach for the construction of Floquet Hamiltonians uses the Fourier transform to map the time-dependent Schrödinger equation to a static eigenvalue equation in an extended Hilbert space~\cite{shirley_solution_1965,sambe_steady_1973}, essentially eliminating the time dependence at the cost of adding an extra dimension to the system. 
Contrary to the unitary case, the (Shirley or Sambe) extended space approach has not been used extensively in the open quantum systems literature. Previous works have shown that this formalism can be extended to the non-unitary case in order to obtain a Floquet Lindbladian acting on an enlarged space~\cite{chen_periodically_2024,szczygielski_floquet_2021,ho_floquetliouville_1986}. The Floquet Lindbladian in Sambe space is, most importantly, of Lindblad form and generates CPTP dynamics~\cite{szczygielski_howland_2020}. This provides a natural starting point for the derivation of effective Floquet Lindbladians.

In this paper we use the Sambe approach to driven-open quantum systems described by a Floquet-Lindblad master equation. For monochromatic drive, where the structure of the Floquet-Lindbladian in Sambe space is tridiagonal, we use a matrix continued fraction method to non-perturbatively resum the Sambe structure and obtain an effective Floquet-Lindbladian acting directly on the physical Hilbert space. We then compute the resolvent, from which the Floquet-Lindblad spectrum and many quantities can be obtained, including frequency-resolved correlation functions in the steady-state. We apply our method to two examples of driven-dissipative quantum systems. First, we consider a two-level system with coherent drive (beyond rotating-wave approximation) and dissipation, for which we compute the spectrum of resonance fluorescence~\cite{mollow1969spectrum,kimble1976theory,walls2008quantum}. Then, we consider a parametrically driven-dissipative quantum dot for which we compute the spectral function and current and discuss dynamical suppression of tunneling. The method we introduce in this work offers a natural approach to Floquet-Lindblad problems beyond the perturbative regime of high-frequency drive and safe from ambiguities due to the non-uniqueness of effective Floquet Lindbladian. It can find natural applications in driven mesoscopic systems, quantum thermodynamics and driven-dissipative many-body systems.

This manuscript is structured as following. In Sec.~\ref{sec:floquet_lindblad} we introduce the framework of Floquet-Linbdlad master equation and its solution with Floquet theory, while in Sec.~\ref{sec:sambe} we review the extended (Sambe) space approach. In Sec.~\ref{sec:cont_fract} we introduce the matrix continued-fraction method to obtain the effective Floquet-Lindbladian, the resolvent and calculate correlation functions in the steady-state. Finally, Sec.~\ref{sec:applications} is devoted to two applications while Sec.~\ref{sec:conclusions} contains our conclusions.

\section{Floquet-Lindblad Master Equation}\label{sec:floquet_lindblad}
We consider the dynamics of an open and periodically driven quantum system whose density matrix $\rho(t)$ at times $t$ evolves according to the time-dependent
Markovian master equation, 
\begin{equation}\label{markov}
   \partial_t\rho(t)=\mathcal{L}(t)\rho(t)=-i\comm{H(t)}{\rho(t)}+\mathcal{D}(t)\rho(t).
\end{equation}
The superoperator $\mathcal{L}$ generating the dynamics is a linear operator acting on the Liouville space $\mathbb{L}=\mathbb{H}\otimes\mathbb{H}^\ast$, where $\mathbb{H}$ is the Hilbert space of the system. We assume it is of Lindblad form and is time-periodic $\mathcal{L}(t+T)=\mathcal{L}(t).$ The evolution consists of a unitary part, determined by a Hamiltonian $H(t),$ and a non-unitary evolution described by the dissipator
\begin{equation}
\mathcal{D}(t)\rho=\sum_i \gamma_i(t)\bqty{L_i(t)\rho L_i^{\dagger}(t)-\frac{1}{2}\acomm{L_i^{\dagger}(t) L_i(t)}{\rho}}.
\end{equation}
We assume rates are always positive $\gamma_i(t)\geq 0$ for all $t\geq 0,$ which guarantees that $\mathcal{L}(t)$ is of Lindblad form for all times and generates completely positive (CP) and trace preserving (TP) dynamics \cite{rivas_quantum_2014,chruscinski_markovianity_2012}. For the time being, we consider the most general case where the time-dependence of $\mathcal{L}(t)$ can arise from either a coherent drive in the Hamiltonian or from time-dependent dissipative rates. 

Introducing the time-evolution superoperator, the evolution of the density matrix from an initial time $t_0$ to a final time $t$ is formally given by $\rho(t)=V(t,t_0)\rho(t_0),$ with the definition
\begin{equation}
    V(t,t_0)=\mathcal{T}e^{\int_{t_0}^t \dd{s}\mathcal{L}(s)}
\end{equation}
where $\mathcal{T}$ indicates time-ordering. The superoperator $V(t,t_0)$ is itself a solution to a linear equation with periodic coefficients $\partial_t V(t,t_0)=\mathcal{L}(t)V(t,t_0)$ with initial condition $V(t_0,t_0)=1.$ Floquet's theorem \cite{floquet_equations_1883} applies, and solutions are of the form 
\begin{equation}
    V(t,t_0)=P(t,t_0)e^{(t-t_0)\mathcal{L}_F}, 
\end{equation}
with $P(t,t_0)$ a periodic superoperator. The difficulty lies in determining the pair $(P,\mathcal{L}_F),$ which would amount to solving the original ordinary differential equation. In equivalence with the unitary case, $\mathcal{L}_F$ is often called a \emph{Floquet Lindbladian}. In principle, we could naively define
\begin{equation}
    \mathcal{L}_F=\frac{1}{T}\log V(T,0).
\end{equation}
However, there is no guarantee that $\mathcal{L}_F$ will be of Lindblad form, even though $\mathcal{L}(t)$ is. 

A question explored in the literature~\cite{hartmann_asymptotic_2017,schnell_is_2020,schnell_high-frequency_2021} is whether it is possible to find an effective \textit{time-independent} Lindbladian $\mathcal{L}_F$, so that $V(T,0)=e^{\mathcal{L}_F T}.$ This differs from the unitary case, where the existence of a Floquet Hamiltonian $\mathcal{H}_F=\frac{i}{T}\log U(T,0)$ is guaranteed because of the Hermitianity of the evolution operator $U$. 

There are three necessary conditions for $\mathcal{L}_F$ to be a valid Lindblad generator: it needs to be (a) trace-preserving, $\Tr(\mathcal{L}_F(A))=\Tr(A)$ for any operator $A$, which guarantees that $\mathcal{L}_F$ has at least one zero eigenvalue corresponding to the steady state, (b) Hermiticity-preserving, $(\mathcal{L}_F(A))^\dagger=\mathcal{L}_F(A^\dagger)$ which implies that the spectrum of $\mathcal{L}_F$ contains complex-conjugated pairs of eigenvalues (it is invariant under complex conjugation, and (c) conditionally completely positive (the evolution $e^{t\mathcal{L}_F}$ it generates is CP)~\cite{wolf_assessing_2008,cubitt2012complexity}. 

The master equation Eq.~(\ref{markov}) can be vectorized, so that operators are mapped to states (represented in a chosen basis as vectors) and superoperators are mapped to operators (represented as matrices)~\cite{Prosen_2008,dzhioev2011,HARBOLA2008191,dorda_auxiliary_2014,arrigoni_master_2018,takahashi96,ojima1981,fazio_manybody_2025}.
This amounts to represent density matrices as pure states $\vert\rho_t \rangle$ in the Liouville space $\mathbb{L}=\mathbb{H}\otimes\mathbb{H}^\ast$ that contains a dual copy of our system. In this formalism the Lindblad master equation takes the form of a Schr\"odinger-like equation 
\begin{align}\label{eqn:master_vector}
\partial_t \vert\rho(t) \rangle=\mathcal{L}(t)\vert\rho(t)\rangle
\end{align}
generated by a non-Hermitian operator $\mathcal{L}(t)$. In the following we use this formalism to describe our approach to Floquet-Lindblad master equation in Eq.~(\ref{markov}), see App.~\ref{app:vectorization} for further details.

\subsection{Floquet theory}\label{sec:floquet}
Since the Lindbladian $\mathcal{L}(t)$ is periodic in time and the evolution of the density matrix is linear we can use Floquet theorem to construct a general solution to Eq.~\eqref{eqn:master_vector}. This takes the form~\cite{szczygielski_application_2014,chen_periodically_2024}
\begin{equation}\label{eq:linearcombi}
    \ket{\rho(t)}=\sum_{j=0}^{(\dim\mathbb{H})^2-1} c_j e^{\mu_j t}\ket{\phi_j(t)}
\end{equation}
where the Floquet states $\ket{\phi_j(t)}$ are periodic in time. Since $\mathcal{L}$ is non-Hermitian, the states $\ket{\phi_j(t)}$ are not necessarily orthogonal. Instead, a biorthogonal basis for the Liouville space $\mathbb{L}=\mathbb{H}\otimes\mathbb{H}^\ast$ can be constructed together with the left states $\bra*{\tilde{\phi}_j(t)}$ \cite{ashida_non-hermitian_2020,brody_biorthogonal_2013}. In this basis, an orthonormality relation exists between left and right states at equal times $\braket*{\tilde{\phi}_j(t)}{\phi_k(t)}=\delta_{jk}.$ Throughout this paper, we assume the completeness of the biorthogonal basis $\mathds{1}=\sum_j \ketbra*{\phi_j(t)}{\tilde{\phi}_j(t)}$, where we have noted $\mathds{1}$ the identity in the Liouville space. The evolution operator can then be decomposed using the left and right Floquet modes
\begin{equation}\label{eq:evolution}
    V(t,t_0)=\sum_j e^{\mu_j(t-t_0)}\ketbra*{\phi_j(t)}{\tilde{\phi}_j(t_0)}.
\end{equation}
In fact, we see that $V(T,0)=\sum_j e^{\mu_j T}\ketbra*{\phi_j(0)}{\tilde{\phi}_j(0)}$ so that the doublet $(\mu_j,\ket{\phi_j(0)})$ can be obtained by diagonalization of the evolution operator over a period. This is, however, computationally demanding because it requires the calculation of the time-ordered exponential $V(T,0)=\mathcal{T}\exp\bqty{\int_0^T \dd{s}\mathcal{L}(s)}.$

\begin{figure*}[t]
    \centering
    \begin{subfigure}{0.4\textwidth}
		\centering
		\includegraphics[height=0.5\textwidth]{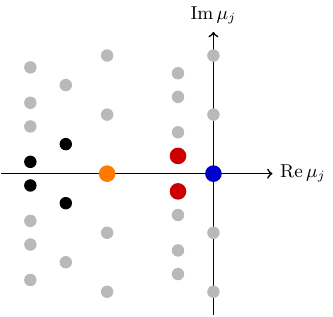}
  		\caption{}
    \label{fig:eigen_schema}
	\end{subfigure}
    \hspace{10pt}
    \begin{subfigure}{0.4\textwidth}
		\centering
		\includegraphics[height=0.5\textwidth]{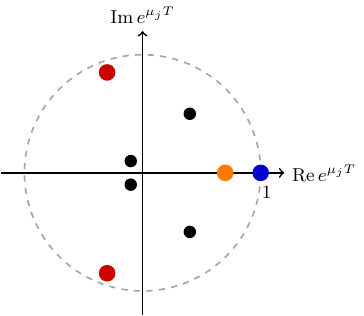}
  		\caption{}
    \label{fig:multipliers_schema}
	\end{subfigure}
\caption{\justifying (a) The spectrum of the Floquet-Lindbladian in the extended space $L_F$ contains eigenvalues $\mu_j$ (colored dots), and their copies along the imaginary axis $\mu_j+in\omega_0, n\in\mathbb{Z}$ (gray dots). (b) The corresponding spectrum of the stroboscopic evolution operator $V(T,0).$ }
\label{fig:spectrum_schema}
\end{figure*}
The Floquet characteristic exponents $\mu_j$, also known as Floquet eigenvalues or Floquet modes, are complex numbers with a negative real part, due to the properties of the Linbdlad superoperator~\cite{chen_periodically_2024} and, in analogy to the quasienergies of non-dissipative Floquet systems, their imaginary part is only defined modulo the driving frequency $\omega_0=2\pi/T.$ The characteristic exponents determine the fate of the solutions at the long-time limit, and the nonequilibrium steady state (which we assume unique) corresponds to the zero eigenvalue which we note with the index $j=0.$ The spectrum of the stroboscopic evolution operator, shown in Fig.~\ref{fig:multipliers_schema} displays the steady-state at $e^{\mu_0 T}=1$ and all the eigenmodes with finite (and negative) real-part lying inside the circle, corresponding to damped modes~\cite{chen_periodically_2024}.

\section{Extended Space Approach}\label{sec:sambe}
Instead of diagonalizing the one-cycle evolution operator $V(T,0)$, we can obtain the Floquet modes by adapting the approach due to J. H. Shirley~\cite{shirley_solution_1965} and H. Sambe~\cite{sambe_steady_1973} to the open quantum system case. The approach consists in using the Fourier transform to map the original time-dependent problem to a time-independent eigenvalue equation in an enlarged space. The corresponding linear operator, the Floquet-Lindblad matrix inherits the properties of the original time-dependent Lindbladian, generating CPTP dynamics when projecting back on the physical space \cite{szczygielski_howland_2020}.\\

Using the Fourier transform
\begin{align}
    \mathcal{L}(t)&=\sum_{m\in\mathbb{Z}} e^{-im\omega_0 t}\mathcal{L}_m\\
    \ket{\phi_j(t)}&=\sum_{m\in\mathbb{Z}} e^{-im\omega_0 t}\ket{\phi_j^m}
\end{align}
where $\omega_0=2\pi/T$ is the drive frequency, we substitute Eq.~(\ref{eq:linearcombi}) into Eq.~(\ref{markov}) and map the time-dependent Lindblad equation to a static eigenvalue problem 
\begin{equation}\label{eqn:recursion}
    \sum_n \pqty{\mathcal{L}_{m-n}+in\omega_0 \delta_{nm}}\ket*{\phi_{j}^n}=\mu_j \ket*{\phi_{j}^m} 
\end{equation}
which, in matrix representation, can be equivalently written as
\begin{widetext}
    \begin{equation}\label{eq:Floquetmatrix}
    \renewcommand*{\arraystretch}{1.3}
       \mqty(\ddots & \vdots & \vdots & \vdots & \\ \cdots & \mathcal{L}_0-i\omega_0 & \mathcal{L}_{-1} & \mathcal{L}_{-2} & \cdots \\ \cdots & \mathcal{L}_1 & \mathcal{L}_0 & \mathcal{L}_{-1} & \cdots \\ \cdots & \mathcal{L}_2 & \mathcal{L}_1 & \mathcal{L}_0+i\omega_0 & \cdots \\  & \vdots & \vdots & \vdots & \ddots)
    \mqty(\vdots \\ \ket*{\phi_{j}^{-1}} \\ \ket*{\phi^0_{j}} \\ \ket*{\phi^1_{j}} \\ \vdots)=\mu_j \mqty(\vdots \\ \ket*{\phi_{j}^{-1}} \\ \ket*{\phi^0_{j}} \\ \ket*{\phi^1_{j}} \\ \vdots)  . 
    \end{equation} 
\end{widetext}
Equation~(\ref{eq:Floquetmatrix}) is equivalent to an eigenvalue problem in an extended Sambe-Liouville space $\mathbb{S}=\mathbb{L}\otimes\mathbb{T},$ with $\mathbb{T}$ being the space of time-periodic functions of period $T,$ 
\begin{equation}
    L_F\ket{\Phi_j}=\mu_j\ket{\Phi_j},
\end{equation}
where the eigenstates are
\begin{equation}
    \ket{\Phi_j}=\sum_m \ket{\phi_j^m}\otimes\ket{m},
\end{equation}
and $L_F$ is the Floquet-Lindblad matrix
\begin{equation}\label{eq:FloquetLindblad}
    L_F=\sum_{mn}\! \pqty{\mathcal{L}_n\otimes\dyad{m\!+\!n}{m}+in\omega_0\delta_{nm}\mathds{1}\otimes\dyad{m}}.
\end{equation}

The spectrum of $L_F$ and the corresponding spectrum of the stroboscopic evolution operator is schematically shown in Fig.~\ref{fig:spectrum_schema}. As we see in panel (a), $L_F$ has a zero eigenvalue corresponding to the steady state $\ket{\Phi_0}.$ The rest of its spectrum consists of complex-conjugated pairs with negative real values. As in the theory of closed Floquet systems, the discrete time symmetry of the original problem leads to a redundancy in the structure of objects in the Sambe space, so that if $\mu_j$ is in the spectrum of $L_F,$ then $\mu_j+in\omega_0$ are also eigenvalues. This is another expression of the fact that the pairs $(\mu_j, \ket{\phi_j(t)})$ and $(\mu_j+in\omega_0,e^{-in\omega_0 t}\ket{\phi_j(t)})$ correspond to the same physical solution of the master equation.

In analogy with the unitary case, the Fourier index in Eq.~\eqref{eq:Floquetmatrix} can be thought of as labeling sites in a fictitious dimension~\cite{ozawa2016synthetic,lin2016photonic,martin2017topological,ozawa2019topological}. Here off-diagonal matrices $\mathcal{L}_{n\neq 0}$ describe the coupling between different sites by absorption ($n>0$) or emission ($n<0$) of $n$ photons, and $\mathcal{L}_{0}$ describes the time-averaged Linbladian over a period of the drive. The key difference with respect to the unitary case is that the model in an effective dimension corresponds to a non-Hermitian lattice, since $\mathcal{L}_{n}$ contains both imaginary (corresponding to coherent evolution) and real part (corresponding to dissipation).

The $in\omega_0$ term corresponds to the conventional linear potential giving rise to the Wannier-Stark ladder \cite{wannier_wave_1960,glueck_wannierstark_2002}, just rotated on the imaginary axis due to the definition of the Lindbladian.

At high-frequency, one expects Wannier-Stark localization for the imaginary part of the eigenvalues. This picture tells us that the real part of the eigenvalues (corresponding to the dissipative dynamics) can also acquire a dependence on the drive frequency even if the dissipator is originally static. This is because at small drive frequencies the ladders are coupled through virtual photon exchange. In the high-frequency regime, one expects Wannier-Stark localization for the imaginary part of the eigenvalues, as the ladders become decoupled and the dissipative dynamics will be the same as for the average motion. 

\section{Matrix-Continued-Fraction Method}\label{sec:cont_fract}
Using Floquet theory, the time-dependent master equation in the Liouville space $\mathbb{L}=\mathbb{H}\otimes\mathbb{H}^*$ is mapped to a static eigenvalue problem in the extended Sambe space $\mathbb{L}\otimes\mathbb{T}.$ The dependence on time is eliminated in exchange with the added dimension of the space of periodic functions $\mathbb{T},$ which is infinite. It is then desirable to be able to interpret the physics in terms of a static and effective Lindbladian that acts on the physical Liouville space.

By focusing on the case where the Floquet-Lindblad matrix $L_F$ is block-tridiagonal, in this Section we introduce a matrix continued fraction method to project onto the zero-photon subspace and obtain an effective Lindbladian acting on the Liouville space $\mathbb{L}$. The obtained effective Lindbladian contains the contribution from higher-harmonics which are resummed into a geometric series. 

\subsection{Effective Floquet-Lindbladian}

In the following we focus on monochromatic drive, such that $\mathcal{L}_{\vert n\vert>2}=0$ and $\mathcal{L}_{\pm 1}\equiv\mathcal{L}_{\pm}$. In this case the resulting tight-binding structure of the Floquet-Lindblad matrix in Eq.~\eqref{eq:Floquetmatrix} allows us to obtain an eigenvalue equation for the zeroth Floquet mode only. The starting point is the recursion relation for the harmonics given in Eq.~(\ref{eqn:recursion}), which for the m-th harmonic now reads
\begin{equation}\label{eq:recursion:m}
    (\mathcal{L}_0\!+\!im\omega_0)\ket*{\phi_j^m}+\mathcal{L}_{-}\ket*{\phi_j^{m+1}}+\mathcal{L}_{+}\ket*{\phi_j^{m-1}}=\mu_j \ket*{\phi_j^m}.
\end{equation}
The Floquet modes can be expressed in a recursive way
\begin{equation}\label{eq:ladder}
\ket{\phi_j^m}=T^{\pm}_m(\mu_j)\mathcal{L}_{\pm}\ket{\phi_j^{m\mp 1}}
\end{equation}
where the matrices $T^{\pm}_m$ serving as raising and lowering operators for the index $m$ admit a recursive definition through matrix continued fractions
\begin{equation}\label{eq:MCFtransfer}
T^{\pm}_m(\mu)=\bqty{\mu-im\omega_0-\mathcal{L}_0-\mathcal{L}_{\mp} T^{\pm}_{m\pm 1}(\mu)\mathcal{L}_{\pm}}^{-1}.
\end{equation}
Substituting Eq.~(\ref{eq:ladder}) into Eq.~(\ref{eq:recursion:m}) and setting $m=0,$ we obtain an eigenvalue equation for the zeroth Floquet mode $\ket{\phi_j^0}$ 
\begin{equation}\label{eq:eigenequation}
    \mathcal{L}_{\mathrm{eff}}(\mu_j)\ket{\phi_j^0}=\mu_j\ket{\phi_j^0}
\end{equation}
where the effective Lindbladian is
\begin{widetext}
\begin{equation}\label{eq:effectivelindblad}
\begin{split}
  \mathcal{L}_{\mathrm{eff}}(\mu)&=\mathcal{L}_0+\mathcal{L}_{-}T^{+}_1(\mu)\mathcal{L}_{+}+\mathcal{L}_{+}T^{-}_{-1}(\mu)\mathcal{L}_{-}=\\
  &=\mathcal{L}_0+\mathcal{L}_{-}\frac{1}{\mu-i\omega_0-\mathcal{L}_0-\mathcal{L}_{-}\frac{1}{\mu-2i\omega_0-\mathcal{L}_0-\dots}\mathcal{L}_{+}}\mathcal{L}_{+}+\mathcal{L}_{+}\frac{1}{\mu+i\omega_0-\mathcal{L}_0-\mathcal{L}_{+}\frac{1}{\mu+2i\omega_0-\mathcal{L}_0-\dots}\mathcal{L}_{-}}\mathcal{L}_{-}.   
\end{split}
\end{equation}
\end{widetext}
We obtain an expression where the average of the Lindbladian over a period of the drive $\mathcal{L}_0$ acquires corrections due to virtual processes that involve an equal number of absorbed and emitted photons. Specifically, the term $\mathcal{L}_{-}T^{+}_1(\mu)\mathcal{L}_{+}$ represents processes where a state in the zero-photon subspace $m=0$ is propagated to all sites of the Floquet chain with $m>0$ by successive absorption of photons of energy $\omega_0$ before returning to $m=0$ by emission of an equal number of photons. The term $\mathcal{L}_{+}T^{-}_{-1}(\mu)\mathcal{L}_{-}$ represents the opposite process where the state first propagates to all $m<0$ before returning to the zero-photon subspace. 
 
The eigenvalues of the effective Lindbladian are found through the secular equation
\begin{equation}\label{eq:secular}
    \det(\mathcal{L}_{\mathrm{eff}}(\mu)-\mu\mathds{1})=0.
\end{equation}
Similar expressions are known for closed Floquet systems \cite{dittrich1998,martinez_floquet_2003,giovannini_floquet_2020,mikami_brillouinwigner_2016,vogl_effective_2020}. As such, the effective Floquet Lindbladian will share the known advantages of its Hamiltonian counterpart over other well-known high-frequency expansions like the Floquet-Magnus and van Vleck expansion: a) unlike the Floquet-Magnus expansion, it has no unphysical dependence on the phase of the drive (since it does not depend on the initial time $t_0$) and b) one can easily obtain all higher-order terms. In fact, Eq.~(\ref{eq:effectivelindblad}) contains contribution of all orders in the frequency, resummed into geometric series. In this sense, $\mathcal{L}_{\mathrm{eff}}$ captures \emph{non-perturbative} effects which are not accessible by high-frequency expansions.

In practice, the effective Lindbladian can be calculated numerically, by terminating the continued fractions for the matrices $T^\pm_{\pm 1}$ at order $N$ with $N>0$ some large integer number. This amounts to setting $T^+_{N+1}=T^-_{-N-1}=0.$ The right choice of $N$ depends on the frequency and amplitude of the drive. In practice $N$ is chosen such that the result does not change by increasing it further. The number of solutions found through Eq.~(\ref{eq:secular}) will also depend on the level of truncation of the continued fractions. Terminating the continued-fractions at order $N$ means that the corresponding secular equation will be a polynomial of order $N$ times the dimension of the original Liouville space $\dim\mathbb{L}.$ 

It is easy to see that $\mathcal{L}_{\mathrm{eff}}$ always has at least one eigenvalue equal to zero. The proof is the following: every harmonic $\mathcal{L}_m$ is trace-preserving \cite{ikeda_nonequilibrium_2021}. In vectorized form, the property of trace preservation is equivalent to the vectorized identity being an element of the left nullspace of every harmonic $\bra{\mathds{1}}\mathcal{L}_m=0.$ By the construction of Eq.~(\ref{eq:effectivelindblad}), we then have $\bra{\mathds{1}}\mathcal{L}_{\mathrm{eff}}=0,$ which implies that $0$ is an eigenvalue. 

Moreover, we show in the Appendix~\ref{app:Leffproperties} that $\mathcal{L}_{\mathrm{eff}}$ is a Hermiticity-preserving operator $\bqty{\mathcal{L}_{\mathrm{eff}}(z)(A)}^\dagger=\mathcal{L}_{\mathrm{eff}}(\overline{z})(A^\dagger)$ so that its spectrum contains complex-conjugated pairs. This property does not depend on the level of truncation of the continued-fractions.

The disadvantage is that the obtained eigenvalue equation becomes self-consistent. This latter point can be avoided by calculating the resolvent operator of the Floquet-Lindblad matrix. The eigenvalues of the Floquet-Lindblad matrix are then identified as poles of the resolvent. Using the continued fraction method, the calculation of the resolvent amounts to inverting a number of matrices of size equivalent to the original Liouville space.

It should, however, be obvious from Eq.~(\ref{eq:effectivelindblad}) and the definition of the raising and lowering matrices Eq.~(\ref{eq:MCFtransfer}) that the self-consistency requirement is no longer needed in the steady state. The steady state corresponds to the eigenvalue $\mu_0=0$ for which we have $\mathcal{L}_{\mathrm{eff}}(0)\ket*{\phi_0^0}=0.$ Knowledge of the nullspace of $\mathcal{L}_{\mathrm{eff}}(0)$ is then sufficient to construct the steady state $\ket{\Phi_0}=\sum_m \ket*{\phi_0^m}\otimes\ket*{m},$ since the harmonics $\ket*{\phi^{m\neq 0}_0}$ can be easily obtained recursively, as described in Eq.~(\ref{eq:ladder}), by multiplying $\ket*{\phi_0^0}$ with a string of matrices.

\subsection{Resolvent operator formalism}
Rather than solving the characteristic polynomial in the extended Sambe space or the self-consistent problem of Eq.~(\ref{eq:secular}) to get the spectrum of the Floquet-Lindblad matrix, we look for the poles of its resolvent operator~\cite{cohen2024atom}. Indeed, if we first generalize Shirley's expression for the unitary evolution operator to the case of open systems
\begin{equation}
    V(t,t_0)=\sum_m e^{-im\omega_0 t}\mel{m}{e^{L_F(t-t_0)}}{0},
\end{equation}
we see that the residue theorem can be used to rewrite the above equation in an integral form
\begin{equation}
   V(t,t_0)=\frac{1}{2\pi i}\sum_m\int \dd{z}e^{z (t-t_0)} \mel{m}{\frac{e^{-im\omega_0 t} }{z- L_F}}{0}.
\end{equation}
Starting with a CPTP map guarantees that the eigenvalues of $L_F$ have a negative real part $\Re\mu_j\leq 0$ and that the integral converges. Such expressions describe the transition amplitude between the initial state at time $t_0$ with zero photons to a final state at time $t$ after absorbing or emitting $m$ photons. This motivates the introduction of the resolvent of $L_F$ defined as
\begin{equation}\label{eq:resolventdef}
    R(z)=\frac{1}{z-L_F}.
\end{equation}
The resolvent operator can be calculated with the matrix continued fraction method. Starting from the definition of the resolvent $\sum_p (z-L_F)_{mp}R_{pn}(z)=\delta_{mn}\mathds{1}$ and taking into account the tridiagonal structure of the linear operator $L_F,$ we get a recursive equation for the resolvent elements
\begin{equation}
(z-im\omega_0-\mathcal{L}_0)R_{mn}-\mathcal{L}_{-}R_{m+1,n}-\mathcal{L}_{+}R_{m-1,n} \!=\!\delta_{mn}\mathds{1}
\end{equation}
From the last recurrence relation we can use the matrix continued fraction method to calculate the elements of the resolvent operator $R.$ In particular, for the diagonal elements, we find
\begin{equation}
R_{mm}(z)\!=\!\bqty{z\!-im\omega_0\!-\mathcal{L}_0\!-\mathcal{L}_{-} T^+_{m+1}\mathcal{L}_{+}\!-\mathcal{L}_{+}T^-_{m-1}\mathcal{L}_{-}}^{-1},
\end{equation}
which for $m=0$ is simply the resolvent of the effective Lindbladian $\mathcal{L}_{\mathrm{eff}}$
\begin{equation}
    R_{00}(z)=\frac{1}{z-\mathcal{L}_{\mathrm{eff}}(z)}.
\end{equation}
For the nondiagonal elements, we find recursive relations
\begin{equation}
    \begin{split}
        R_{mn}(z)=\begin{cases}
            T^+_m(z) \mathcal{L}_+ R_{m-1,n}(z) \quad  &m>n \\
            T^-_m(z)\mathcal{L}_- R_{m+1,n}(z) \quad  &m<n
        \end{cases}
    \end{split}
\end{equation}
which imply that, by iteration, nondiagonal elements can be expressed as a function of the diagonal elements times a string of matrices.
\begin{equation}
 R_{mn}=
    \begin{cases}
        T^+_m \mathcal{L}_+ T^+_{m-1}\mathcal{L}_+\dots T^+_{n+1}\mathcal{L}_+  R_{nn} &\qif m>n\\
        T^-_m \mathcal{L}_- T^-_{m+1}\mathcal{L}_-\dots T^-_{n-1}\mathcal{L}_-  R_{nn} &\qif m<n
    \end{cases}
\end{equation}

\subsection{Micromotion} 
We have assumed that particular solutions of the master equation are of the form $\ket*{\rho_j(t)}=e^{\mu_j t}\ket*{\phi_j(t)}.$ Using the effective Floquet-Lindbladian one can find its eigenvectors in the `zero photon' subspace $\ket*{\phi_j^0}.$ Then, the continued fraction method can be used to recursively construct all other modes $\ket*{\phi_j^m}$ through Eq.~(\ref{eq:ladder}). As a result, the Floquet states $\ket*{\phi_j(t)}$ are 
\begin{equation}
\begin{split}
    \ket*{\phi_j(t)}&=\sum_m e^{-im\omega_0 t}\ket*{\phi_j^m}\\
    &=\ket*{\phi_j^0}+e^{-i\omega_0 t}T^+_1\mathcal{L}_+ \ket*{\phi_j^0}+e^{i\omega_0 t}T^-_{-1}\mathcal{L}_- \ket*{\phi_j^0}+\dots\\
    &\equiv K_j(t)\ket*{\phi_j^0}.
\end{split}
\end{equation}
The solutions to the master equation are then given by
\begin{equation}
    \ket*{\rho_j(t)}=e^{\mu_j t}K_j(t)\ket*{\phi_j^0}.
\end{equation}
The operator $K_j(t)$ could therefore be associated to the micromotion, while $\ket*{\phi_j^0}$ are Floquet modes that represent average motion since
\begin{equation}
    \ket*{\phi_j^0}=\frac{1}{T}\int_0^T \dd{t}\ket{\phi_j(t)}.
\end{equation}
To make the connection more explicit, we see that the evolution operator given in Eq.~(\ref{eq:evolution}) can now be rewritten as
\begin{equation}
    V(t,t_0)=\sum_j e^{\mu_j(t-t_0)}K_j(t)\dyad*{\phi_j^0}{\tilde{\phi}_j^0}\tilde{K}_j(t_0),
\end{equation}
where $\tilde{K}_j(t)$ is obtained by applying the continued fraction method to the left recursion problem $\bra*{\tilde{\phi}_j^m}(\mathcal{L}_0+im\omega_0)+\bra*{\tilde{\phi}_j^{m+1}}\mathcal{L}_{-}+\bra*{\tilde{\phi}_j^{m-1}}\mathcal{L}_{+}=\mu_j \bra*{\tilde{\phi}_j^{m}}.$ $K_j(t)$ and $\tilde{K}_j(t)$ are by construction time-periodic operators. Their respective Fourier harmonics are
\begin{equation}
\begin{split}
    K_j^m         &=T^\pm_m \mathcal{L}_\pm T^\pm_{m\mp 1}\mathcal{L}_\pm\dots T^\pm_{\pm 1}\mathcal{L}_{\pm}\\
    \tilde{K}_j^m &=\mathcal{L}_{\pm}T^\pm_{\pm 1}\dots\mathcal{L}_\pm T^\pm_{m\mp 1}\mathcal{L}_\pm T^\pm_m
\end{split}  
\end{equation}
and $K_j^0=\tilde{K}_j^0=1.$ In order for the biorthogonality of the Floquet states to hold, the following condition must be respected
\begin{equation}
     \braket*{\tilde{\phi}_i(t)}{\phi_j(t)}=\mel*{\tilde{\phi}_i^0}{\tilde{K}_i(t)K_j(t)}{\phi_j^0}=\delta_{ij}.
\end{equation}

\subsection{Building the steady state}\label{sec:steadystate}
In the steady state, the nonlinearity of $\mathcal{L}_\mathrm{eff}$ is removed and the eigenvalue problem Eq.~(\ref{eq:eigenequation}) becomes $\mathcal{L}_{\mathrm{eff}}(0)\ket*{\phi_0^0}=0.$ The effective Lindbladian becomes a constant matrix
\begin{equation}
  \mathcal{L}_{\mathrm{eff}}(0)=  \mathcal{L}_0+\mathcal{L}_{-}T^{+}_1(0)\mathcal{L}_{+}+\mathcal{L}_{+}T^{-}_{-1}(0)\mathcal{L}_{-}.
\end{equation}
Once $\ket*{\phi_0^0}$ is found, the rest of the harmonics are easily obtained by multiplying the zeroth harmonic with the operator $K_0^m,$ $\ket*{\phi_0^m}=K_0^m\ket*{\phi_0^0}.$ Crucially, $K_0^m$ is now also simply given by a string of constant operators. The steady state in the extended space is therefore 
\begin{equation}
    \ket{\Phi_0}=\sum_m K_0^m\ket{\phi_0^0}\otimes \ket{m} 
\end{equation}
or, equivalently, in the physical space,
\begin{equation}
    \ket{\phi_0(t)}=K_0(t)\ket{\phi_0^0}.
\end{equation}
The left eigenvector $\bra*{\tilde{\Phi}_0}$ can be built in a similar manner.

\subsection{Calculation of correlation functions}\label{sec:correlation}
Multitime correlation functions can be calculated by making use of the quantum regression theorem \cite{gardinerzoller,breuer}. For $A,B$ some system operators, their correlation function  $C(t_1,t_2)=\expval{A(t_1)B(t_2)}$ is given by
\begin{equation}
     C(t_1,t_2)=\mel{\mathds{1}}{A V(t_1,t_2)B V(t_2,t_0)}{\rho(t_0)}
\end{equation}
which holds for $t_1\geq t_2\geq t_0.$ Generalizing the calculation in \cite{arrigoni_master_2018} to the time-periodic case, we will show that correlation functions in the steady state can be written in a way analogous to the Lehmann spectral representation for closed systems~\cite{Scarlatella_2019}.
As a first step, we introduce the relative time $t_r=t_1-t_2$ and the average time $t_a=(t_1+t_2)/2.$ In the steady state, $t_0\to\infty,$ correlation functions inherit the discrete time translation invariance of the Lindbladian, so that $C(t_1+T,t_2+T)=C(t_1,t_2),$ consequently they can be expanded in a Fourier series in the average time $C(t_1,t_2)=\sum_m e^{-im\omega_0 t_a}C_m(t_r).$ The Fourier transform of the components $C_m(t_r)$ with respect to the relative time
\begin{equation}
    C_m(\omega)=\int_{-\infty}^{\infty}\!\dd{t_r} e^{i\omega t_r} \frac{1}{T}\int_{-T/2}^{T/2}\!\dd{t_a} e^{im\omega_0 t_a} C(t_1,t_2)
\end{equation}
is called the Wigner transformation of the two-time correlation function $C(t_1,t_2)$ \cite{tsuji_correlated_2008}. Once the calculation is performed (see Appendix~\ref{app:Floquet correlations} for details), it is convenient to return to the Floquet representation. The two representations are in fact related through the following relation
\begin{equation}\label{eq:representations}
    C_{mn}(\omega)=C_{m-n}(\omega+\frac{m+n}{2}\omega_0).
\end{equation}
In the Floquet representation, we show that correlation functions depend on the resolvent operator $(\omega-iL_F)^{-1}$. In the steady state, we find that two-time correlation functions can be expressed compactly in the extended space as
\begin{equation}\label{eq:correlationfunctions}
     C_{mn}(\omega)=\mel*{\tilde{\Phi}_0}{(A\!\otimes \!\mathscr{T}_{-m}) \frac{i}{\omega-iL_F} (B\!\otimes\! \mathscr{T}_n)}{\Phi_0},
\end{equation}
where $\ket{\Phi_0}$ and $\bra*{\tilde{\Phi}_0}$ are, respectively, the right and left zero eigenvectors of $L_F$ and $\mathscr{T}$ is a translation operator which acts on Floquet indices by shifting them in Fourier space
$\mathscr{T}_m\equiv\sum_n \dyad{n+m}{n}.$ Operators $A$ and $B$ are acting on the Liouville space corresponding to the density matrix $\rho.$ The frequency $\omega$ is taken to be real. Convergence is then guaranteed if the eigenvalues of $L_F$ have negative real parts $\Re(\mu_j)<0.$ A finite shift $\omega+i0^+$ needs to be added so that eigenvalues are slightly shifted away from the imaginary axis. 

The case where the operators are not time-ordered $C'(t_1,t_2)=\expval{B(t_2)A(t_1)}, t_1\geq t_2,$ is treated by rewriting the correlation function in terms of its complex conjugate $\expval{B(t_2)A(t_1)}=\expval{A^\dagger(t_1)B^\dagger(t_2)}^\ast.$ With a similar calculation, we obtain 
\begin{equation}
     C'_{mn}(\omega)=\mel*{\tilde{\Phi}_0}{(A^\dagger\!\otimes\! \mathscr{T}_{m}) \frac{i}{\omega+iL_F} (B^\dagger\!\otimes\! \mathscr{T}_{-n})}{\Phi_0}.
\end{equation}

\section{Applications} \label{sec:applications}

In this Section we present two applications of the Sambe approach and matrix-continued-fraction method to two periodically driven and dissipative quantum systems. First we consider a two-level system which is coherently and periodically driven by an external field and experiences static decoherence. In the second application we consider a fermionic quantum dot which is parametrically driven by a periodic in time gate voltage (or energy shift) and coupled to pump and loss dissipation. In both cases we discuss the structure of the Floquet-Lindblad spectrum and characterize the steady-state by looking at observables and correlations functions.

\subsection{Periodically driven two-level system}
We start by studying a two-level system (TLS) described by a time-periodic Hamiltonian and static dissipation

\begin{equation}
    \mathcal{L}(t)\rho=-i\comm{H(t)}{\rho}+\gamma\left(\sigma_-\rho\sigma_+-\frac{1}{2}\acomm{\sigma_+\sigma_-}{\rho}\right)
\end{equation}

with a Hamiltonian describing the TLS in a linearly polarized field
\begin{equation}\label{eqn:Htls}
    H(t)=\frac{\Delta}{2}\sigma_z+A\cos(\omega_0 t-\phi)\sigma_x
\end{equation}
and the dissipator describing spontaneous emission of strength $\gamma$. The $\sigma_i$ are the standard Pauli matrices for $i=x,y,z$ and the raising or lowering operators for $i=\pm$. In Eq.~(\ref{eqn:Htls}) $A,\omega_0,\phi$ are respectively the amplitude, frequency and phase of the drive, while $\Delta$ is the energy scale of the TLS ground-to-excited state transition. This model in its simplicity captures the main effects of Floquet drive and dissipation. It was considered in Ref.~\cite{schnell_is_2020} to discuss the existence of a Floquet-Lindbladian obtained from the stroboscopic evolution.

\begin{figure}[t]
    \centering
		\includegraphics[width=0.4\textwidth]{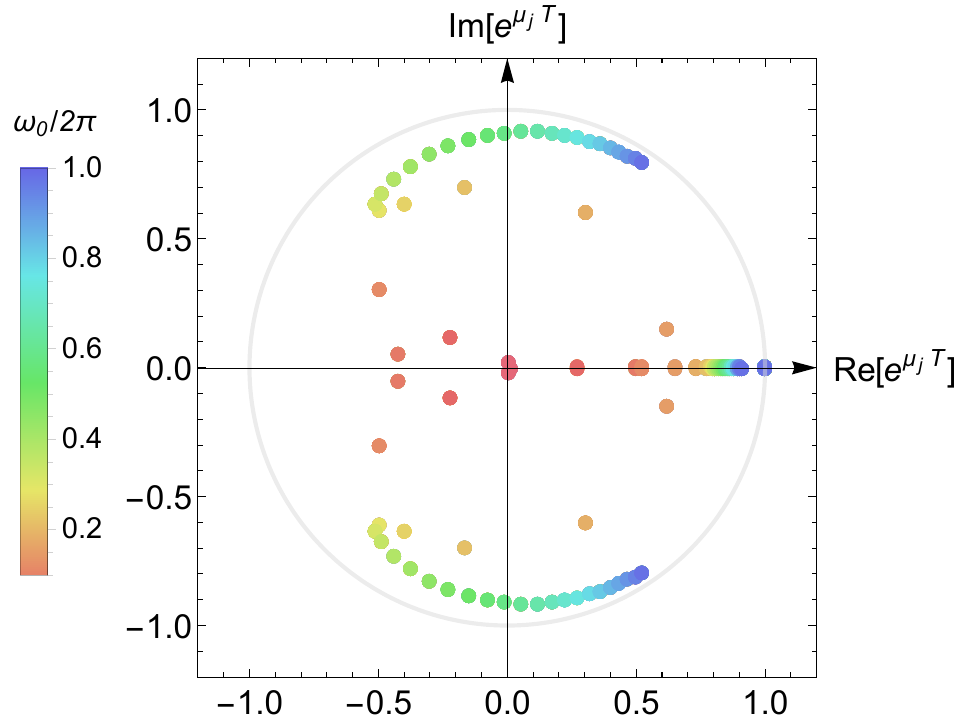}
\caption{\justifying Spectrum of the stroboscopic evolution operator as obtained from diagonalization of the Floquet-Lindblad matrix $L_F$ at different drive frequencies $\omega_0,$ for $A=\Delta,\gamma=0.1\Delta,\Delta=1, \phi=0.$ }
    \label{fig:stroboscopic}
\end{figure}
First, we calculate the eigenvalues $\mu_j$ of the Floquet-Lindblad matrix $L_F$ defined in Eq.~(\ref{eq:FloquetLindblad}), through exact diagonalization, after truncation where we restrict the Fourier indices $m\in\mathbb{Z}$ with a cutoff $\abs{m}\leq 30.$ In Fig.~\ref{fig:stroboscopic}, we plot the corresponding eigenvalues $e^{\mu_j T}$ of the stroboscopic evolution operator $V(t_0+T,t_0)$ in the complex plane and as the frequency of the drive is swept in the region $\omega_0/\Delta\in \bqty{0.1, 2\pi}.$ We fix the rest of the model parameters to $\Delta=1, A=\Delta,\gamma=0.1\Delta,\phi=0.$ 

We see that all eigenvalues $e^{\mu_j T}$ stay within the unit circle (indicated with a gray line) and come in complex-conjugated pairs, as one would expect from a CPTP map~\cite{chen_periodically_2024}. As the frequency is increased, eigenvalues tend to approach a few points inside the unit circle (marked in blue). These points correspond to  the high-frequency limit $A/\omega_0\ll 1,$ where the spectrum of $L_F$ is that of the time-averaged Lindbladian $\mathrm{spec}(\mathcal{L}_0)=\{0,-i\Delta-\gamma/2,i\Delta-\gamma/2,-\gamma\}$ modulo integer multiples of $i\omega_0.$

\begin{figure}[t]
    \centering
        \includegraphics[width=0.35\textwidth]{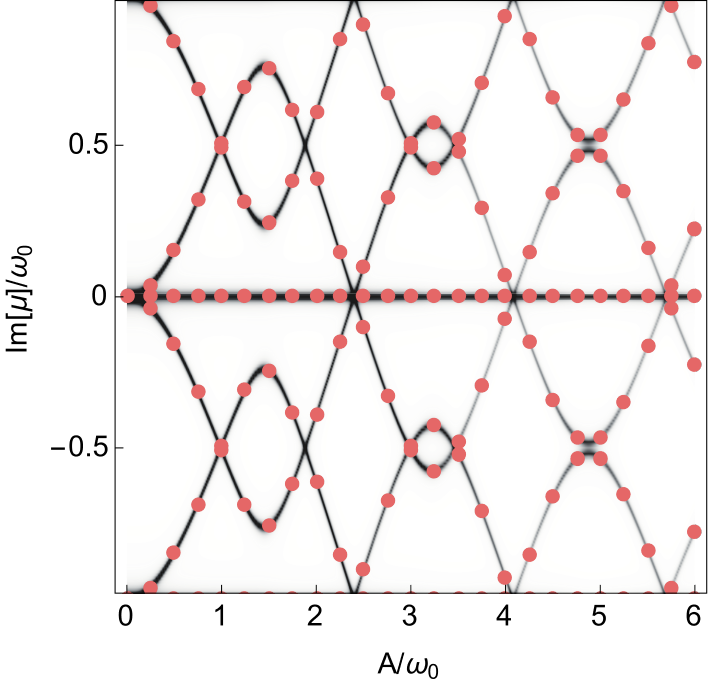}
\caption{\justifying Comparison between the Floquet-Lindblad spectrum obtained from the exact diagonalization of the evolution operator (red dots) and the matrix-continued fraction (black lines). Parameters are chosen as $\omega_0=\Delta/2,\gamma=0.01\Delta,\Delta=1, \phi=0.$}
    \label{fig:TLS_spec}
\end{figure}

In Fig.~\ref{fig:TLS_spec} we show the agreement between the imaginary part of the eigenvalues obtained through exact diagonalization of $L_F$ (red dots) and the continued fraction method (black lines). The plot demonstrates that the MCF method can be used in the strong coupling regime $A/\omega_0\gtrsim 1$ where high-frequency expansions typically fail. 
The spectrum is obtained graphically by calculating the imaginary part of the component $i\Tr R_{00}(-i\omega)$ since, as shown in App.~\ref{app:resolvent}, this quantity is a sum of Lorentzians centered around the imaginary part of the eigenvalues $\Im(\mu_j)$ with width given by the real part $-\Re(\mu_j).$ We only plot a few points obtained through exact diagonalization since the result coincides with the one obtained from the MCF method. 

\begin{figure*}[t]
    \centering
    \begin{subfigure}{0.3\textwidth}
		\centering
		\includegraphics[height=\textwidth]{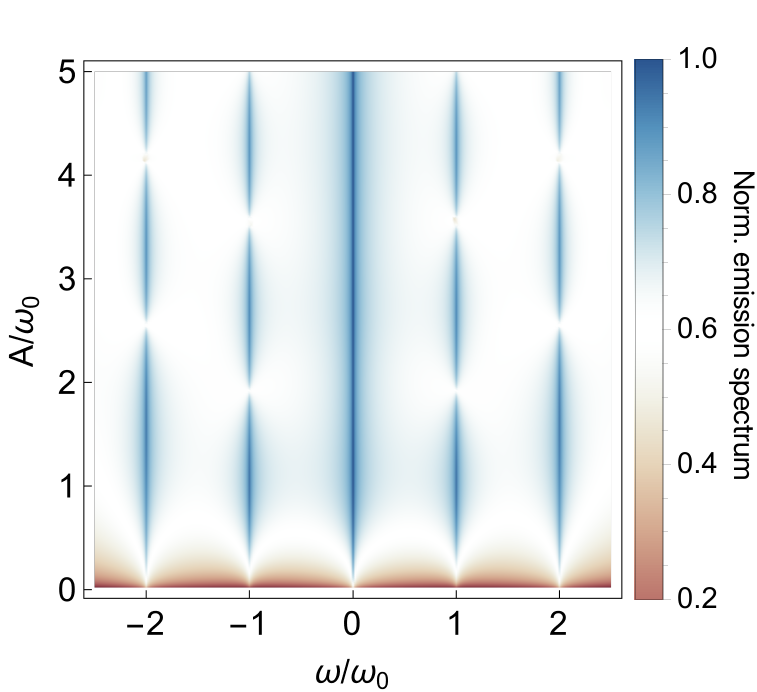}
  		\caption{}
    \label{fig:emission:a}
	\end{subfigure}
    \hfill
    \begin{subfigure}{0.28\textwidth}
		\centering
		\includegraphics[height=\textwidth]{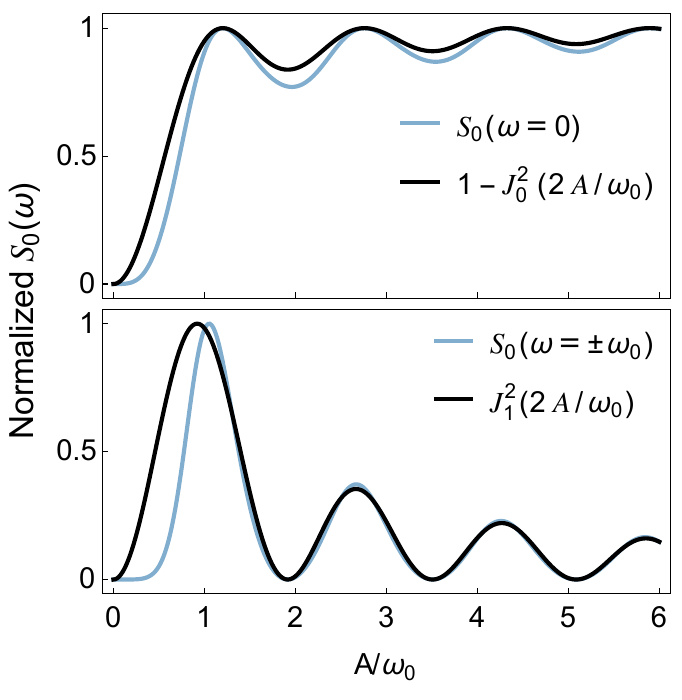}
  		\caption{}
    \label{fig:emission:b}
	\end{subfigure}
    \hfill
   \begin{subfigure}{0.3\textwidth}
		\centering
		\includegraphics[height=\textwidth]{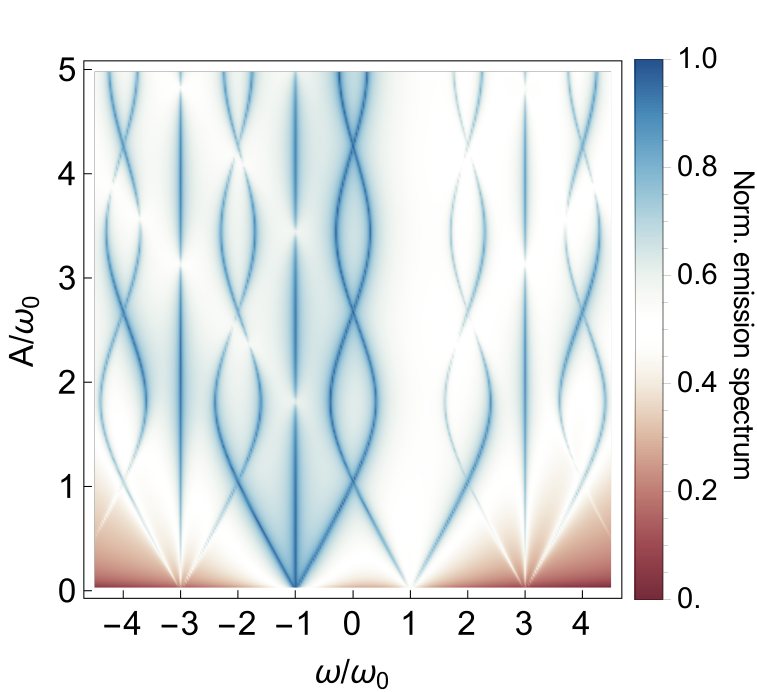}
  		\caption{}
    \label{fig:emission:c}
	\end{subfigure}
\caption{\justifying (a),(c): Emission spectra of the periodically driven TLS as a function of the emission frequency $\omega$ and the amplitude of the drive $A$ in units of the drive frequency $\omega_0$. The intensity of the emission is normalized to $1$. In (a) the level splitting is zero $\Delta=0,$ while in (c) the drive frequency is resonant to the splitting $\Delta=\omega_0.$ (b): Comparison of the intensity of the emission lines at $\omega=0$ (top panel) and at $\omega=\pm\omega_0$ (bottom panel) with Bessel functions of the first kind $J^2_p(2A/\omega_0)$ in the case $\Delta=0$. The rest of the parameters are $\omega_0=1,\gamma=0.01,\phi=0.$}
\label{fig:emission}
\end{figure*}

We then calculate the two-time correlation function $S(t_1,t_2)=\expval{\sigma_+(t_1)\sigma_-(t_2)}$, which encodes the properties of the emission spectrum. In the rotating wave approximation regime where the amplitude of the drive is considered to be small, this quantity reduces to the well studied Mollow triplet of resonance fluorescence~\cite{mollow1969spectrum,kimble1976theory,walls2008quantum}. Here we will discuss how the emission evolves in a non-perturbative regime of drive amplitudes.

In the Wigner representation discussed in Sec.~\eqref{sec:correlation}, we calculate the time average over one period and take its Fourier transform with respect to the relative time to obtain
\begin{equation}
    S_0(\omega)=2\Re \int_{0}^{+\infty}\dd{t_{r}}e^{i\omega t_{r}} \frac{1}{T}\int_{-T/2}^{+T/2} \dd{t_{a}}S(t_1,t_2).
\end{equation}
In the Floquet representation $S_0(\omega)=S_{00}(\omega)$, where we have used Eq.~(\ref{eq:representations}). From Eq.~(\ref{eq:correlationfunctions}) we get
\begin{equation}\label{eq:RFspectrum}
\begin{split}
      S_{0}(\omega) &=2\Re \mel*{\tilde{\Phi}_0}{(\sigma_+\otimes I)\frac{i}{\omega-iL_F}(\sigma_-\otimes I)}{\Phi_0}\\
      &=2\Re\sum_{mn}\mel*{\tilde{\phi}_0^m}{\sigma_+ R_{mn}(-i\omega)\sigma_-}{\phi_0^n}\\
&=-2\Im\sum_{mnp}\sum_{j}\frac{\mel*{\tilde{\phi}_0^m}{\sigma_+}{\phi_j^{m+p}}\mel*{\tilde{\phi}_j^{n+p}}{\sigma_-}{\phi_0^{n}}}{\omega-p\omega_0-i\mu_j}
\end{split}
\end{equation}
where $I=\sum_m \dyad{m}$ is the identity in the space $\mathbb{T}$ and the Pauli matrices have been written in vectorized representation $\sigma_\pm\ket{\rho}=\ket{\sigma_\pm\rho}$. In Eq.~(\ref{eq:RFspectrum}), the second equation involving the resolvent is used for the numerical calculation, while the last equation is used for the interpretation of the results. 

Indeed, from the last equation we see that there are two contributions to the emission spectrum. The first contribution, corresponding to $j=0$ and the zero eigenvalue $\mu_0=0$, is a coherent component comprised of delta peaks at multiples of the drive frequency
\begin{equation}
\begin{split}
 S_\mathrm{coh}(\omega)&=2\pi\sum_p\delta(\omega-p\omega_0)\\ &\times\sum_{mn}\mel*{\tilde{\phi}_0^m}{\sigma_+}{\phi_0^{m+p}}\mel*{\tilde{\phi}_0^{n+p}}{\sigma_-}{\phi_0^{n}}.   
\end{split}
\end{equation}
The coherent component is the Fourier transform of the term $\expval{\sigma_+(t_1)}\expval{\sigma_-(t_2)}$ after it has been time-averaged.

The second component $j\neq 0$ is an incoherent contribution corresponding to inelastic scattering of light
\begin{equation}
S_\mathrm{inc}(\omega)=-2\Im\sum_{mnp}\sum_{j\neq 0}\frac{\mel*{\tilde{\phi}_0^m}{\sigma_+}{\phi_j^{m+p}}\mel*{\tilde{\phi}_j^{n+p}}{\sigma_-}{\phi_0^{n}}}{\omega-p\omega_0-i\mu_j}.
\end{equation}

Writing the eigenvalues of the Lindbladian as $\mu_j=-i\omega_j-\gamma_j,$ with $\omega_j,\gamma_j\in\mathbb{R},$ we see that the emission spectrum is comprised of copies of three Lorentzian functions centered at $\omega=\omega_j+p\omega_0,$ each having width $\gamma_j.$ The weight of the peaks is given by the matrix elements $\mel*{\tilde{\phi}_0^m}{\sigma_+}{\phi_j^{m-p}}\mel*{\tilde{\phi}_j^{n-p}}{\sigma_-}{\phi_0^{n}}$ which describe transitions between different Floquet states. The three peaks are the Mollow triplet, dressed by the Floquet drive.

The results of the numerical calculations are shown on Fig.~\ref{fig:emission} where we plot the emission spectrum $S_0(\omega)$ as a function of $\omega/\omega_0$ and $A/\omega_0$. We focus on two cases, first considering the splitting $\Delta=0$ and then considering the case where the frequency of the drive is on resonance with the splitting $\Delta=\omega_0.$

In the first case (Fig.~\ref{fig:emission:a}), when the level splitting is zero $\Delta=0,$ resonances appear at multiples of the drive frequency $\omega=p\omega_0.$ As the drive amplitude increases, the emission at fixed frequency $\omega$ has an oscillating behavior, proportional to Bessel functions of the first kind $J^2_p(2A/\omega_0).$ This is illustrated in Fig.~\ref{fig:emission:b} where the emission spectrum (with intensity normalized so that it takes values between $0$ and $1$) is plotted as a function of $A/\omega_0.$ We find that the emission lines at $\omega=p\omega_0$ are suppressed at specific values which correspond to the zeroes of the Bessel functions $J_p(2A/\omega_0).$ Only the central peak at $\omega=0$ does not exhibit this destructive interference, and shows constructive interference from all channels $p\neq 0.$ As a result of the Bessel sum rule $\sum_{p} J^2_p(2A/\omega_0)=1,$ the weight of the central peak is proportional to $\sum_{p\neq 0} J^2_p(2A/\omega_0)=1-J^2_0(2A/\omega_0).$

In the second case (Fig.~\ref{fig:emission:c}) we choose the driving frequency to be on resonance with the splitting between the two levels, $\Delta=\omega_0.$ In the rotating wave approximation regime where the amplitude of the drive is considered to be small, we see the appearance of the Mollow triplet~\cite{mollow1969spectrum,kimble1976theory,walls2008quantum} centered around $\omega=-\omega_0.$ Replicas of the Mollow triplet appear at odd multiples of the driving frequency in agreement with previous works~\cite{browne_resonance_2000,yan_resonance_2016}. Beyond this regime, the emission exhibits a more complicated behavior involving Bessel function oscillations as well as an asymmetry between the intensities at positive and negative frequencies. As the amplitude is increased, the weight of emission lines at positive frequencies is suppressed, and the fluorescence spectrum is dominated by peaks at negative frequencies. 

\begin{figure}[t]
    \centering
		\includegraphics[height=0.35\textwidth]{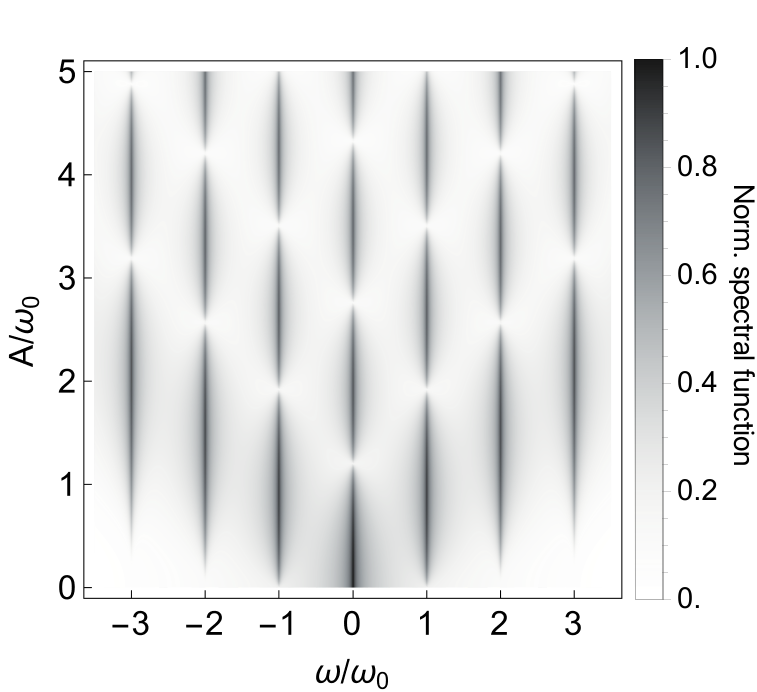}
\caption{\justifying Spectral function $\mathcal{A}(\omega)$ of the driven resonant level model as a function of the amplitude of the drive $A$ and the frequency $\omega$ in units of the drive frequency $\omega_0$ obtained through the continued fraction method. The rest of the parameters are set to $\gamma_L=\gamma_R=0.01, \varepsilon_d=0.$}
 \label{fig:RLM:dos}
\end{figure}

\subsection{Periodically driven quantum dot}\label{sec:RLM}
In this section we consider a periodically driven quantum dot coupled to two fermionic leads. We assume the quantum dot can be approximated by a single resonant level and that a voltage gate is applied to modulate the energy level sinusoidally. We write the Hamiltonian of the quantum dot
\begin{equation}
   H(t)=\varepsilon_d(t) d^\dagger d=(\varepsilon_d+2A\cos\omega_0 t)d^\dagger d
\end{equation}
where $d^\dagger,d$ are fermionic creation and annihilation operators, $\varepsilon_d$ is the position of the unperturbed electronic level, and $A,\omega_0$ are the amplitude and frequency of the drive, respectively. We assume that the influence of the leads is described by the dissipator 
\begin{equation}\label{eq:RLMdissipator}
\mathcal{D}\rho=\sum_\alpha \pqty{\gamma_\alpha^+\mathcal{D}\bqty{d^\dagger}+\gamma_\alpha^-\mathcal{D}\bqty{d}}\rho
\end{equation}
where the rates $\gamma_\alpha^+=\gamma_\alpha f_\alpha(\varepsilon_d)$ and $\gamma_\alpha^-=\gamma_\alpha (1-f_\alpha(\varepsilon_d))$ depend on the coupling $\gamma_\alpha$ of the dot to the reservoir $\alpha=L,R,$ and the Fermi-Dirac distribution at chemical potential $\mu_\alpha$ and temperature $T_\alpha,$ $f_a(\omega)=(e^{(\omega-\mu_\alpha)/T_\alpha}+1)^{-1.}$ We use the notation $\mathcal{D}\bqty{L}\rho=L\rho L^\dagger-\frac{1}{2}\acomm{L^\dagger L}{\rho}.$ It will also be useful to define the total coupling to the leads $\gamma\equiv\sum_\alpha \gamma_\alpha.$

Expanding into Fourier series $\mathcal{L}(t)=\sum_m \mathcal{L}_m e^{-im\omega_0 t},$ the zeroeth harmonic corresponds to a static dot coupled to two reservoirs,  $\mathcal{L}_0\bqty{\bullet}=-i\varepsilon_d\comm{d^\dagger d}{\bullet}+\mathcal{D}\bqty{\bullet},$ while for $m=\pm 1,$ $\mathcal{L}_{\pm}\bqty{\bullet}=-iA\comm{d^\dagger d}{\bullet}.$ 

We adopt the superfermion representation \cite{dzhioev2011} in order to vectorize the Lindblad master equation. In this approach, one doubles the original Hilbert space $\mathbb{H}$ by introducing a `tilde' copy $\tilde{\mathbb{H}}$. The density matrix $\rho$ becomes a vector $\ket{\rho}$ in $\mathbb{H}\otimes \tilde{\mathbb{H}}.$ Further details are provided in the Appendix~\ref{app:vectorization}. In vectorized form, the harmonics of the Lindbladian are
\begin{equation}
\begin{split}
\mathcal{L}_0=&-i\varepsilon_d(N-\tilde{N})-\frac{\sum_\alpha\gamma_\alpha^- -\gamma_\alpha^+}{2}(N+\tilde{N})\\ &-i\sum_\alpha \gamma_\alpha^- d \tilde{d}-\sum_\alpha\gamma_\alpha^+(1-id^\dagger\tilde{d}^\dagger)  \\
\mathcal{L}_{\pm}=&-iA(N-\tilde{N})
\end{split}
\end{equation}
where tilde fermionic operators act in $\tilde{\mathbb{H}}.$ $N$ and $\tilde{N}$ are the number operators in the original and the tilde space, respectively. Working in the composite Fock space spanned by states $\ket*{M,\tilde{M}}=\ket{M}\otimes\ket*{\tilde{M}},$ the matrices $\mathcal{L}_0,\mathcal{L}_\pm$ are matrices of size 4. In the extended space approach, they become the blocks of the Floquet-Lindblad matrix $L_F.$

The problem can be simplified by observing that $L_F$ commutes with the operator $\mathcal{K}=i(N-\tilde{N}),$ since the physical and tilde fermions are always created or annihilated simultaneously, and their difference is conserved. As a result, when the Lindbladian is acting on a Fock state of the doubled Hilbert space $\ket*{M,\tilde{N}},$ it preserves the number $K=M-\tilde{N}.$ It can therefore be block-diagonalized in terms of blocks of the number $K.$ For the present case, $K$ has only three possible values: $K=-1,0,1.$ Finally, the steady state can be found in the $K=0$ sector since if $L_F\ket{\Phi_{0}}=0,$ then $\mathcal{K}\ket{\Phi_{0}}=0$ follows \cite{dorda_auxiliary_2014}. 

The steady state for this example is particularly simple because the $K=0$ sector is block-diagonal in the Sambe space, $L_F^{K=0}=\oplus_m (\mathcal{L}^{K=0}_0+im\omega_0).$ The corresponding eigenvalues are therefore the same as of a static dot coupled to two reservoirs modulo integer multiples of $i\omega_0,$ $\mu_j=\{0+im\omega_0,-\gamma +im\omega_0\}.$ The steady state corresponding to the eigenvalue $\mu_0=0$ is easily found by solving $(\mathcal{L}^{K=0}_0+im\omega_0)\ket{\phi_0^m}=0.$ Since $(\mathcal{L}^{K=0}_0+im\omega_0)$ is invertible for $m\neq 0,$ we find that the only harmonic in the steady state is the zeroth $\ket{\Phi_0}=\sum_m\ket{\phi_0^m}\otimes\ket{m}\delta_{m0}.$ 

Defining the Green's function of the dot $G^r(t_1,t_2)= -i\theta(t_1-t_2)\expval{\acomm{d(t_1)}{d^\dagger(t_2)}},$ and using Eq.~(\ref{eq:correlationfunctions}) we calculate the time-averaged spectral function $\mathcal{A}(\omega)=-\frac{1}{\pi}\Im G^r_{0}(\omega+i\eta).$ In Appendix~\ref{app:Floquet correlations} we show that the time-averaged spectral function is given by
\begin{equation}
\begin{split}
     \mathcal{A}(\omega)=&-\frac{1}{\pi}\Im\mel*{\tilde{\Phi}_0}{d \frac{1}{\omega-i L_F}d^\dagger}{\Phi_0}\\
     &-\frac{1}{\pi}\Im\mel*{\tilde{\Phi}_0}{d^\dagger \frac{1}{\omega+i L_F}d}{\Phi_0}^\ast.
\end{split}
\end{equation}
Identifying $(\omega\pm i L_F)^{-1}=\pm i R(\pm i\omega),$ the spectral function can be calculated through the matrix continued fraction method. For the present problem, the steady state has only an $m=0$ component so that only the diagonal element $R_{00}$ is needed for the calculation.
\begin{figure}[t]
		\centering
		\includegraphics[width=0.4\textwidth]{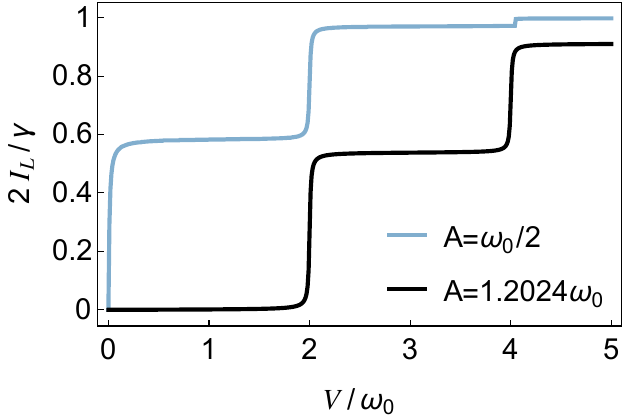}
\caption{\justifying Current-voltage characteristics for different driving amplitudes showing the suppression of current at the Bessel function zero of $J_0(2A/\omega_0)$. $\omega_0=1, \gamma_L=\gamma_R=0.01, \varepsilon_d=0.$}
 \label{fig:RLM:current}
\end{figure}
The result of the numerical calculation is shown in Fig.~\ref{fig:RLM:dos} where the spectral function is plotted as a function of $\omega/\omega_0$ and $A/\omega_0.$ The maxima of the spectral function $\mathcal{A}(\omega)$ are shown with black. We fix the frequency of the drive at $\omega_0=1$ and the energy level of the dot $\varepsilon_d=0.$ From the analysis of the spectrum of $L_F$ in the steady state we expect the peaks of the spectral function to be at $\omega=m\omega_0.$ Moreover, we observe that the spectral function at energy $\omega=m\omega_0$ shows destruction of tunneling at the zeroes of the Bessel function of the first kind $J_m(2A/\omega_0).$ This is similar to the unitary case, however the width of the resonances is controlled by the total coupling to the leads $\gamma.$ In fact, we show in Appendix~\ref{app:exactRLM} through an exact calculation that the time-averaged retarded Green's function in the steady-state is given by
\begin{equation}\label{eq:BesselGreen}
    G^r_0(\omega)=\sum_m \frac{J^2_m\pqty{\frac{2A}{\omega_0}}}{\omega-\varepsilon_d-m\omega_0+i\frac{\gamma}{2}}.
\end{equation}
This is a well-known result, usually derived through non equilibrium Green's functions formalism and serves as a benchmark for our method.

The destruction of tunneling is more apparent when looking at the current passing through the dot. In the Appendix~\ref{app:exactRLM} we calculate the steady-state time-averaged current from the left lead to the resonant level, using the Meir-Wingreen expression connecting the current with the Green's function of the dot~\cite{jauho_timedependent_1994,PLATERO20041}
\begin{equation}
 \expval{I_L}=\sum_{m}\int\frac{\dd{\omega}}{2\pi}J^2_m\pqty{\frac{2A}{\omega_0}}T(\omega\!-\!m\omega_0)\bqty{f_L(\omega)-f_R(\omega)}  
\end{equation}
where $T(\omega)=\frac{\gamma_L\gamma_R}{(\omega-\varepsilon_d)^2+\gamma^2/4}$ is the transmission coefficient through a resonant level at energy $\varepsilon_d.$
At the zeroes of $J_m(2A/\omega_0)$ the current is suppressed, as shown in Fig. \ref{fig:RLM:current} where the current is suppressed at the first zero of the Bessel function $J_0(2A/\omega_0).$

\section{Conclusions} \label{sec:conclusions}

In this work we have introduced a method to study driven and dissipative quantum systems described by a Floquet-Lindblad master equation with a periodic in time Lindblad superoperator. In contrast with recent literature which has focused on the question of defining an effective Lindbladian in an appropriate high-frequency limit, here we have used Floquet Theory and the Shirley-Sambe representation to map the time-dependent problem to a static one in the extended Sambe space of harmonics of the drive. The advantage of this representation is that it naturally allows to introduce a Floquet-Lindblad operator which describes a well defined CPTP map, yet in an enlarged (infinite-dimensional) space. 

We then consider specifically the case of monochromatic drive where the structure in Sambe space is block-tridiagonal, allowing in practice an exact solution of the problem via a continued-fraction method. In particular, an effective Floquet-Lindbladian acting only on the physical (doubled) Hilbert space is obtained, from which spectrum and stationary state are obtained. In practice we show that the natural object to manipulate in this framework is the resolvent, which satisfies simple recursive equations. We discuss finally how to reconstruct the full micromotion as well as correlation functions on the steady-state.

As applications of the method we consider two paradigmatic models of driven-dissipative quantum systems, a dissipative two-level system in a coherent time-periodic field and a parametrically driven dot with pump and losses. In the former case we demonstrate the agreement between continued fraction method and exact diagonalization in Sambe space and discuss the resonance fluorescence spectrum (Mollow triplet) away from the textbook limit of weak drive and rotating-wave approximation. In the latter, we discuss the physics of dynamical suppression of tunneling.

The method introduced in this work can be extended to a variety of driven-dissipative Floquet-Lindblad problems, including bosonic models relevant for example to driven superconducting circuits~\cite{floquetqubit}, models displaying dissipative phase transitions, such as the boundary-time crystal~\cite{iemini2018boundary}, whose fate in presence of Floquet drive remains a largely open question, as well as models which are gaussian and translational invariant for which the structure of the Sambe space can be tackled exactly in momentum space leading to an efficient numerical solution. This could be the case for examples of models relevant for quantum transport in presence of local dissipation~\cite{damanet2019controlling,visuri2023nonlinear} as well as models of driven-dissipative bosons~\cite{PhysRevB.111.165131}. Here a comparison with different approaches such as Keldysh field theory could provide further insights on the method.
These examples represent fruitful research directions to be explored in the coming future.

\begin{acknowledgments}
We acknowledge financial support from the ERC consolidator grant No.~101002955 - CONQUER.
\end{acknowledgments}

\appendix

\section{Vectorization of the Lindblad master equation}\label{app:vectorization}
In order to transform the Lindblad equation into a Schr\"odinger-like equation, we need a way to vectorize the density matrix. The vectorization will result in rewriting the master equation as
\begin{equation}
    \dv{t}\ket{\rho(t)}=\mathcal{L}\ket{\rho(t)}
\end{equation}
where now $\mathcal{L}$ is to be understood as a non-Hermitian matrix. The left eigenvectors $\bra{l_a}$ and right eigenvectors $\ket{r_a}$ of this matrix, defined as
\begin{equation}
    \begin{split}
        \mathcal{L}\ket{r_a}&=\lambda_a\ket{r_a}\\
        \bra{l_a}\mathcal{L}&=\bra{l_a}\lambda_a
    \end{split}
\end{equation}
can be used to construct a biorthogonal basis \cite{brody_biorthogonal_2013,ashida_non-hermitian_2020}. In such a basis the identity can be represented by $I=\sum_a \ket{r_a}\bra{l_a},$ expressing the completeness of the basis. An orthonormality relation exists between left and right eigenvectors $\braket{l_a}{r_b}=\delta_{ab},$ but not necessarily between the left and right eigenvectors themselves $\braket{l_a}{l_b}\neq \delta_{ab},\braket{r_a}{r_b}\neq \delta_{ab}.$  In the biorthogonal basis, the vectorized density matrix can be written as \cite{fazio_manybody_2025}
\begin{equation}
\begin{split}
    \ket{\rho(t)}&=e^{\mathcal{L}t}\ket{\rho(0)} = \ket{r_0}+\sum_{a\neq 0}e^{\lambda_a t}\ket{r_a}\braket{l_a}{\rho(0)}\\
     &=\ket{\rho_{ss}}+\sum_{a\neq 0}c_a e^{\lambda_a t}\ket{r_a}.
\end{split}
\end{equation}
In the last line the steady-state $\ket{\rho_{ss}}$ (which, for simplicity, we assume is unique) is identified with the right eigenvector $\ket{r_0}$ corresponding to the eigenvalue $\lambda_0=0.$ All the other eigenvalues have a negative real part (decay modes) and come in complex conjugate pairs. 

An important role in this framework is played by the vectorized identity $\ket{\mathds{1}},$ also called the `left vacuum vector'. In vectorized form, $\Tr(\mathcal{L}\rho)=\mel{\mathds{1}}{\mathcal{L}}{\rho}.$ From the trace preserving property of the Lindbladian we conclude that 
\begin{equation}
  \forall \rho,\quad  \mel{\mathds{1}}{\mathcal{L}}{\rho}=0\Rightarrow \bra{\mathds{1}}\mathcal{L}=0,
\end{equation}
which implies that $\bra{\mathds{1}}$ can be identified to the left-eigenvector of $\mathcal{L}$ with eigenvalue $\lambda_0=0, \bra{\mathds{1}}=\bra{l_0}.$ From these definitions, it follows that only the steady-state is tracefull since $\braket{l_0}{r_0}=1\Rightarrow \Tr(r_0)=\Tr(\rho_{ss})=1,$ while all other solutions are traceless $\Tr(r_{a\neq 0})=0.$

\subsection{Superfermion representation}
One vectorization approach is the superfermion representation \cite{dzhioev2011}, also presented in \cite{dorda_auxiliary_2014,arrigoni_master_2018}. The density matrix $\rho(t)$ is an operator in a Fock space spanned by states $\ket*{M},$ $I=\sum \dyad*{M}, \braket*{N}{M}=\delta_{NM}.$ One then makes a copy of the initial Fock space, a so-called `tilde' Fock space where $\tilde{I}=\sum \dyad*{\tilde{M}}, \braket*{\tilde{N}}{\tilde{M}}=\delta_{NM}.$ The left vacuum in the composite Fock space is defined as
\begin{equation}
    \ket{\mathds{1}}=\sum_M \ket*{M}\otimes\ket*{\tilde{M}}.
\end{equation}
Any operator, such as the density matrix, can then be vectorized
\begin{equation}
   \ket{\rho(t)}=(\rho(t)\otimes \tilde{I})\ket{\mathds{1}}=\sum_{MN} \rho_{NM}(t) \ket*{M}\otimes\ket*{\tilde{N}},
\end{equation}
with $\rho_{NM}(t)=\mel*{M}{\rho(t)}{N}$. From $\Tr\rho(t)=1$ it follows that $\braket{\mathds{1}}{\rho(t)}=1.$ In a similar fashion, operator expectation values can be written as matrix elements
\begin{equation}
    \expval{A(t)}=\Tr(A\rho(t))=\mel{\mathds{1}}{A\rho(t)}{\mathds{1}}=\mel{\mathds{1}}{A}{\rho(t)}.
\end{equation}
In order to vectorize the Lindbladian we need the following tilde conjugation rules for fermionic operators
\begin{equation}
    \begin{split}
        &d_i\ket{\mathds{1}}=-i\tilde{d}^{\dagger}_i\ket{\mathds{1}}\\ 
        &d^{\dagger}_i\ket{\mathds{1}}=-i\tilde{d}_i\ket{\mathds{1}}. 
    \end{split}
\end{equation}
The same relations hold for any operator which is a linear combination of $d_i,d_i^\dagger$ with real coefficients. Using the tilde conjugation rules and the fact that the density matrix commutes with tilde operators, one can vectorize the Lindblad master equation $\mathcal{L}\bqty{\rho}\to \mathcal{\hat{L}}\ket{\rho}$ \cite{arrigoni_master_2018}
\begin{equation}
    \mathcal{L}\bqty{\rho}=-i\comm{H}{\rho}+\sum_{ij} \gamma_{ij} L_i\rho L_j^\dagger-\frac{1}{2}\acomm{L^\dagger_j L_i}{\rho}
\end{equation}
Assuming the jump operators are linear combinations of $d_i$ and $d_i^\dagger$ and that $H$ is Hermitian and contains an even number of $d_i,d_i^\dagger,$ we get the vectorized Lindbladian
\begin{equation}
\begin{split}
       \mathcal{\hat{L}}&=-i(H-\tilde{H})\\ &+\sum_{ij} \gamma_{ij}\pqty{-i L_i \tilde{L}_j-\frac{1}{2}L^\dagger_j L_i-\frac{1}{2}\tilde{L}_i^\dagger \tilde{L}_j}. 
\end{split}
\end{equation}

\section{Properties of the effective Lindbladian}\label{app:Leffproperties}
In this section we establish some properties of the effective Floquet Lindbladian
\begin{equation}\label{app:eq:effectivelindblad}
 \mathcal{L}_{\mathrm{eff}}(\mu)=\mathcal{L}_0+\mathcal{L}_{-}T^+_{1}(\mu)\mathcal{L}_{+}+\mathcal{L}_{+}T^-_{-1}(\mu)\mathcal{L}_{-}.
\end{equation}
To this end, it is useful to realize that the continued fraction expression of the raising and lowering operators $T^\pm_m$ admits a diagrammatic representation
\begin{equation}\label{app:eq:diagrammatic}
\begin{split}
    T^\pm_m(z)&=\frac{1}{z\!-\!im\omega_0\!-\!\mathcal{L}_0\!-\!\mathcal{L}_{\mp} T^{\pm}_{m\pm 1}\mathcal{L}_{\pm}}\\ &= R^0_{m}+R^0_m\mathcal{L}_\mp R^0_{m\pm 1}\mathcal{L}_\pm R^0_m \\&+R^0_m\mathcal{L}_\mp R^0_{m\pm 1}\mathcal{L}_\mp R^0_{m\pm 2}\mathcal{L}_\pm R^0_{m\pm 1}\mathcal{L}_\pm R^0_m \\&+\dots   
\end{split}
\end{equation}
where we have defined
\begin{equation}\label{app:eq:R0}
    R^0_m(z)=\frac{1}{z-im\omega_0-\mathcal{L}_0}.
\end{equation}
After setting boundary sites $\pm N$ of the Floquet chain,  we see that such an expansion of $T^\pm_m$ resums all loops that start and end at position $m$ after having visited all sites $m'>m$ (in the case of $T^+_m$) or $m'<m$ (in the case of $T^-_m$). Equation~(\ref{app:eq:effectivelindblad}) can be interpreted as a resummation of all such diagrams. 
\subsection{Trace preservation} 
For a time periodic Lindbladian 
\begin{equation}
\mathcal{L}(t)\bullet=-i\comm{H(t)}{\bullet}+\sum_i \gamma_i(t)\bqty{L_i\bullet L_i^{\dagger}-\frac{1}{2}\acomm{L_i^{\dagger}L_i}{\bullet}}
\end{equation}
every harmonic is trace preserving:
\begin{equation}
\begin{split}
    \Tr(\mathcal{L}_m\rho)&=-i\Tr(\comm{H_m}{\rho})\\&+\sum_i \gamma_i^m \Tr(L_i\rho L_i^{\dagger}-\frac{1}{2}L_i^{\dagger}L_i\rho-\frac{1}{2}\rho L_i^{\dagger}L_i)=0 
\end{split}
\end{equation}
where the cyclicity of the trace was used. In vectorized form the trace preservation is equivalent to the vectorized identity being an element of the left nullspace of every harmonic $\bra{\mathds{1}}\mathcal{L}_m=0.$ By construction, we then have $\bra{\mathds{1}}\mathcal{L}_{\mathrm{eff}}=0.$ In particular, this implies that the effective Lindbladian always has at least one eigenvalue equal to zero.\\

\subsection{Complex-conjugated spectrum}
The time-periodic Lindbladian superoperator $\mathcal{L}_t$ has the Hermiticity-preserving property $(\mathcal{L}_t(A))^\dagger=\mathcal{L}_t(A^\dagger)$ for any operator $A.$ This implies that the harmonics of the Lindbladian obey the relation
\begin{equation}
    (\mathcal{L}_{m}(A))^\dagger=\mathcal{L}_{-m}(A^\dagger).
\end{equation}
From this we conclude that only the zero-harmonic is Hermiticity-preserving. Moreover, it is easy to check that superoperator products where harmonics $\mathcal{L}_{m}$ and $\mathcal{L}_{-m}$ come in pairs satisfy the following relation
\begin{equation}
   (\mathcal{L}_{m}\mathcal{L}_{-m}(A))^\dagger=\mathcal{L}_{-m}\mathcal{L}_{m}(A^\dagger)  
\end{equation}
so that terms which are symmetric under exchange of indices are also Hermiticity-preserving, e.g. a term like $\mathcal{L}_{m}\mathcal{L}_{-m}+\mathcal{L}_{-m}\mathcal{L}_{m},$ or $\mathcal{L}_{m}\mathcal{L}_0\mathcal{L}_{-m}+\mathcal{L}_{-m}\mathcal{L}_0\mathcal{L}_{m}$ preserves Hermiticity.

As an intermediate step to checking whether $\mathcal{L}_{\mathrm{eff}}$ is Hermiticity-preserving, we need to check a property of the resolvent $R^0_m(z),$ defined in Eq.~(\ref{app:eq:R0}), namely 
\begin{equation}\label{app:eq:R0herm}
    (R^0_m(z)(A))^\dagger=R^0_{-m}(\overline{z})(A^\dagger).
\end{equation}
As a result, and using the perturbative expansion (\ref{app:eq:diagrammatic}) we see that
\begin{equation}
    (T^\pm_m(z)(A))^\dagger=T^\mp_{-m}(\overline{z})(A^\dagger)
\end{equation}
which means that 
\begin{equation}
     (\mathcal{L}_{m}T^\pm_n(z)\mathcal{L}_{-m}(A))^\dagger=\mathcal{L}_{-m}T^\mp_{-n}(\overline{z})\mathcal{L}_{m}(A^\dagger).
\end{equation}
Finally, we have that the effective Lindbladian 
$\mathcal{L}_{\mathrm{eff}}(z)=\mathcal{L}_0 +\mathcal{L}_{-}T^+_1(z)\mathcal{L}_{+}+\mathcal{L}_{+}T^-_{-1}(z)\mathcal{L}_{-}$ is Hermiticity-preserving
\begin{equation}
\begin{split}
   \bqty{\mathcal{L}_{\mathrm{eff}}(z)(A)}^\dagger&= \bqty{(\mathcal{L}_0 +\mathcal{L}_{-}T^+_1(z)\mathcal{L}_{+}+\mathcal{L}_{+}T^-_{-1}(z)\mathcal{L}_{-})(A)}^\dagger\\&=(\mathcal{L}_0 +\mathcal{L}_{+}T^-_{-1}(\overline{z})\mathcal{L}_{-}+\mathcal{L}_{-}T^+_{1}(\overline{z})\mathcal{L}_{+})(A^\dagger)\\
   &=\mathcal{L}_{\mathrm{eff}}(\overline{z})(A^\dagger). 
\end{split}
\end{equation}
This means that if the pair $(\mu_j,\phi_j)$ is an eigensolution,  so is the pair $(\mu_j^\ast,\phi_j^\dagger),$ implying that the spectrum of the effective Lindbladian consists of pairs of complex-conjugated eigenvalues. Moreover, the spectrum retains this property at any level of truncation of the continued-fractions, since terminating the continued fraction for the raising and lowering operators at any level does not alter the crucial property which is that the Lindbladian harmonics $\mathcal{L}_{\pm}$ always appear in pairs.

\section{Properties of the resolvent operator}\label{app:resolvent}
By inspection we can establish an important property of the linear operator $L_F$
\begin{equation}\label{eq:covariance}
    L_F(\mu_j)_{m+p,n+p}=L_F(\mu_j-ip\omega_0)_{m,n}.
\end{equation}
Using Eq. (\ref{eq:covariance}) we can also establish an equivalent relation for the resolvent
\begin{equation}\label{eq:resolvent:covariance}
    R(z-i p\omega_0)_{m,n}=R(z)_{m+p,n+p}.
\end{equation}
Using the completeness relation of the Floquet replicas $\ket{\Phi_{j,p}}=e^{ip\omega_0 t} \ket{\Phi_j},$ the identity in the Sambe-Liouville space is $1_S=\sum_{j,p}\dyad*{\tilde{\Phi}_{j,p}}{\Phi_{j,p}}$, and we can write the resolvent as
\begin{equation}\label{eq:resolvent}
    R_{mn}(z)=\sum_{j,p} \frac{\dyad*{\phi_j^{m-p}}{\tilde{\phi}_j^{n-p}}}{z-i p\omega_0-\mu_j}.
\end{equation}
This form of the resolvent will appear when calculating correlation functions. For the numerical evaluation of this object it is more convenient to use the continued fraction method.

Writing the eigenvalues as $\mu_j=-i\omega_j-\gamma_j$ with $\omega_j,\gamma_j$ real we see that
\begin{equation}
    i\Tr R_{00}(-i\omega)=-\sum_{j,p} \frac{\Tr(\dyad*{\phi_j^p}{\tilde{\phi}_j^p})}{\omega-(\omega_j+p\omega_0)+i\gamma_j}
\end{equation}
where $\omega$ is on the real axis and the trace is taken in the Liouville space. The imaginary part of this expression is therefore a sum of Lorentzians centered around $\omega=\omega_j=-\Im\mu_j,$ with width $\gamma_j=-\Re\mu_j.$

\section{Spectral representation of Floquet-Green's functions}\label{app:Floquet correlations}
In this section we show how to use the quantum regression theorem to calculate two-times correlation functions between two system operators $C(t_1,t_2)=\expval{A(t_1)B(t_2)}$ in the steady-state and obtain a spectral representation of the retarded Floquet-Green's function of an open quantum system.

A first object to look at is the expectation value of an operator $\expval{A(t)}=\Tr(A \rho(t)).$ Assuming the left $\bra*{\tilde{\phi}_j}$ and right $\ket*{\phi_j}$ states form a biorthogonal basis at all times, the identity operator in Liouville space is
\begin{equation}
    \mathds{1}=\sum_j \dyad*{\phi_j(t)}{\tilde{\phi}_j(t)}.
\end{equation}
The left and right eigenvectors are orthogonal $\braket*{\tilde{\phi}_j(t)}{\phi_k(t)}=\delta_{jk}$ only for equal time arguments. For the average of an operator we find
\begin{equation}
\begin{split}
\expval{A(t)} &=\Tr(A \rho(t))=\braket{\mathds{1}}{A\rho(t)}\\&=\sum_j \braket{\mathds{1}}{\phi_j(t)}\braket*{\tilde{\phi}_j(t)}{A\rho(t)}\\&=\mel*{\tilde{\phi}_0(t)}{A}{\rho(t)}=\sum_j c_j e^{\mu_j t} A_{0j}(t)
\end{split}
\end{equation}
where we use that only the mode $j=0$ is tracefull \cite{chen_periodically_2024} so that $\braket{\mathds{1}}{\phi_j(t)}=\delta_{j0},$ and we introduced the shorthand $A_{jk}(t)=\mel*{\tilde{\phi}_j(t)}{A}{\phi_{k}(t)}.$ Since the $\phi$ states are time-periodic we can Fourier expand them, so that
\begin{equation}
    A_{jk}(t)=\sum_p e^{-ip\omega_0 t} \alpha_{jk}^{p}
\end{equation}
with the identification of the Fourier coefficients as
\begin{equation}\label{eq:Fouriercoef}
     \alpha_{jk}^{p}\equiv \sum_m \mel*{\tilde{\phi}_j^m}{A}{\phi_{k}^{m+p}}.
\end{equation}
Using the definition of the evolution operator in terms of the Floquet modes $V(t,t')=\sum_j e^{\mu_j(t-t')}\ketbra*{\phi_j(t)}{\tilde{\phi}_j(t')}$
we verify that
\begin{equation}
      \ket{\rho(t)}=V(t,0)\ket{\rho(0)}=\sum_j c_j e^{\mu_j t}\ket{\phi_j(t)}
\end{equation}
with $c_j=\braket*{\tilde{\phi}_j(0)}{\rho(0)},$ but also,
\begin{equation}
\begin{split}
   \ket{\rho(t)}&=V(t,t_0)\ket{\rho(t_0)}\\&=\sum_j e^{\mu_j (t-t_0)}\ket{\phi_j(t)}\braket*{\tilde{\phi}_j(t_0)}{\rho(t_0)}
   \end{split}
\end{equation}
which implies that $\braket*{\tilde{\phi}_j(t_0)}{\rho(t_0)}=c_j e^{\mu_j t_0}.$\\

Using the quantum regression theorem and the definition of the non-unitary evolution operator, we calculate the two-times correlation function $\expval{A(t_1)B(t_2)}$ for $t_1>t_2$
\begin{equation}
\begin{split}
       C(t_1,t_2)&=\mel{\mathds{1}}{A V(t_1,t_2)B V(t_2,t_0)}{\rho(t_0)}\\
       &=\sum_{jk} e^{\mu_j(t_1-t_2)}e^{\mu_k(t_2-t_0)}\mel{\mathds{1}}{A}{\phi_j(t_1)}\\&\times\mel*{\tilde{\phi}_j(t_2)}{B}{\phi_k(t_2)}\braket*{\tilde{\phi}_k(t_0)}{\rho(t_0)}\\
       &=\sum_{jk} c_k e^{\mu_j(t_1-t_2)}e^{\mu_k t_2} A_{0j}(t_1)B_{jk}(t_2).
\end{split}
\end{equation} 
We note that the result does not depend on the choice of initial time $t_0.$ In the steady-state $k=0$ and
\begin{equation}
\begin{split}
   C^\infty(t_1,t_2)&=\sum_{j} e^{\mu_j(t_1-t_2)} A_{0j}(t_1)B_{j0}(t_2)\\&=\sum_j\sum_{mn} e^{\mu_j(t_1-t_2)} e^{-im\omega_0 t_1}e^{-in\omega_0 t_2} \alpha^m_{0j}\beta^n_{j0}.  
\end{split}
\end{equation} 
Working in the Wigner representation \cite{uhrig_positivity_2019,tsuji_correlated_2008}, we introduce the average and relative times
\begin{equation}
    \begin{split}
        t_r&=t_1-t_2\\
        t_a&=(t_1+t_2)/2
    \end{split}
\end{equation}
so that $t_{1,2}=t_{a}\pm t_{r}/2.$ We can then calculate the Wigner representation of any two-time correlation function by taking the double integral
\begin{equation}
     C_l(\omega)=\int\dd{t_{r}}\frac{1}{T}\int_{-T/2}^{+T/2} \dd{t_{a}}e^{i\omega t_{r}+il\omega_0 t_{a}} C(t_1,t_2).
\end{equation}

\begin{widetext}
The Wigner transformation of $\theta(t_1-t_2) C^\infty(t_1,t_2)$ is
\begin{equation}
\begin{split}
   C_l(\omega) &= \sum_{j}\sum_{mn}\int_{0}^{\infty}\dd{t_{r}} e^{i\omega t_{r}}\frac{1}{T}\int_{-T/2}^{T/2} \dd{t_{a}}e^{il\omega_0 t_{a}}e^{-im\omega_0 t_1-in\omega_0 t_2} \alpha^m_{0j}\beta^n_{j0}\\
   &=\sum_{j}\sum_{mn}\int_{0}^{\infty}\dd{t_{r}} e^{\bqty{i\omega+\mu_j-i(\frac{m-n}{2})\omega_0} t_{r}}\frac{1}{T}\int_{-T/2}^{T/2} \dd{t_{a}}e^{i(l-m-n)\omega_0 t_{a}}\alpha^m_{0j}\beta^n_{j0}\\
     &=  \sum_{j,m}\int_{0}^{\infty}\dd{t_{r}} e^{\bqty{i\omega+\mu_j-i(m-\frac{l}{2})\omega_0} t_{r}}\alpha^m_{0j}\beta^{l-m}_{j0}.
\end{split}
\end{equation}
\end{widetext}
Using the Laplace transform of the exponential we get 
\begin{equation}
    C_{l}(\omega)=\sum_{j,m}\frac{i\alpha^m_{0j}\beta^{l-m}_{j0}}{\omega-(m-\frac{l}{2})\omega_0-i\mu_j}.
\end{equation}
The integral converges if $\Re(\mu_j)<0.$ A finite shift $\omega+i0^+$ needs to be added so that eigenvalues are slightly shifted away from the imaginary axis. It is often useful to be able to pass from the Wigner representation to the Floquet matrix representation, which can be done with use of the following relation \cite{tsuji_correlated_2008}
\begin{equation}
    C_{kl}(\omega)=C_{k-l}(\omega+\frac{k+l}{2}\omega_0).
\end{equation}

In the present case,
\begin{equation}
\begin{split}
     C_{kl}(\omega)&=\sum_{j,m}\frac{i \alpha^m_{0j}\beta^{k-l-m}_{j0}}{\omega+(k-m)\omega_0-i\mu_j}\\&=\sum_{j,p}\frac{i \alpha^{k-p}_{0j}\beta^{p-l}_{j0}}{\omega+p\omega_0-i\mu_j}.
\end{split}
\end{equation}
This expression relates the correlation function with the resolvent operator of the Floquet matrix in Sambe space $R(z)=(z-L_F)^{-1}$. Using the definition of the Fourier coefficients (\ref{eq:Fouriercoef}), we rewrite
\begin{equation}
\begin{split}
     C_{kl}(\omega)&=\sum_{mn}\sum_{j,p}\frac{i\mel*{\tilde{\phi}_0^m}{A}{\phi_j^{m+k-p}}\mel*{\tilde{\phi}_j^{n+l-p}}{B}{\phi_0^{n}} }{\omega+p\omega_0-i\mu_j}\\
     &=i\sum_{mn}\mel*{\tilde{\phi}_0^{m-k}}{A \bqty{\frac{1}{\omega-iL_F}}_{mn}B}{\phi_0^{n-l}}\\
     &=i\mel*{\tilde{\Phi}_0}{(A\otimes\mathscr{T}_{-k}) \frac{1}{\omega-iL_F}(B\otimes\mathscr{T}_l)}{\Phi_0}.
\end{split}
\end{equation}
Equivalently, we can write $ C_{kl}(\omega)$ in terms of the resolvent
\begin{equation}
\begin{split}
     C_{kl}(\omega)&\sum_{mn}\mel*{\tilde{\phi}_0^{m-k}}{A R_{mn}(-i\omega)B}{\phi_0^{n-l}}\\
     &=\mel*{\tilde{\Phi}_0}{(A\otimes\mathscr{T}_{-k}) R(-i\omega) (B\otimes\mathscr{T}_l)}{\Phi_0}.
\end{split}
\end{equation}
We have introduced the translation operator $\mathscr{T}$ which acts on Floquet indices by shifting them in Fourier space $\mathscr{T}_n\equiv\sum_p \dyad{p+n}{p}$ so that $\mathscr{T}_n\ket{m}=\ket{m+n}$ and $\bra{m}\mathscr{T}_n=\bra{m-n}$.

Defining the retarded Green's function as $G^r(t_1,t_2)=-i\theta(t_1-t_2)\expval{\acomm{A(t_1)}{B(t_2)}}$ and performing a similar calculation for the term $\expval{B(t_2)A(t_1)}=\expval{A^\dagger(t_1)B^\dagger(t_2)}^\ast$ leads to 
\begin{equation}
\begin{split}
    G^r_{kl}&=\mel*{\tilde{\Phi}_0}{(A\otimes\mathscr{T}_{-k}) \frac{1}{\omega-i L_F}(B\otimes\mathscr{T}_l)}{\Phi_0}\\&+\mel*{\tilde{\Phi}_0}{(A^\dagger\otimes \mathscr{T}_{k}) \frac{1}{\omega+i L_F}(B^\dagger\otimes\mathscr{T}_{-l})}{\Phi_0}^\ast. 
\end{split}
\end{equation}

For $A=d$ and $B=d^\dagger$ in particular we find
\begin{equation}
\begin{split}
     G^r_{kl}(\omega)&=\mel*{\tilde{\Phi}_0}{d\otimes\mathscr{T}_{-k} \frac{1}{\omega-i L_F} d^\dagger\otimes\mathscr{T}_l}{\Phi_0}\\&+\mel*{\tilde{\Phi}_0}{d^\dagger \otimes\mathscr{T}_{k} \frac{1}{\omega+i L_F}d\otimes\mathscr{T}_{-l}}{\Phi_0}^\ast.
\end{split}
\end{equation}
A special role is played by the diagonal component $k=l=0$ whose imaginary part corresponds to the average of the spectral function over one period 
\begin{equation}
\begin{split}
      A(\omega)=&-\frac{1}{\pi}\Im G^r_{l=0}(\omega)\\=&-\frac{1}{\pi}\Im\bqty{\mel*{\tilde{\Phi}_0}{d \frac{1}{\omega-i L_F}d^\dagger}{\Phi_0}}\\&-\frac{1}{\pi}\Im\bqty{\mel*{\tilde{\Phi}_0}{d^\dagger \frac{1}{\omega+i L_F}d}{\Phi_0}^\ast}.
\end{split}
\end{equation}

\section{Exact dynamics of the driven and dissipative resonant level model}\label{app:exactRLM}
For the resonant level model with periodic drive and dissipation described by Hamiltonian $H(t)=\pqty{\varepsilon_d+2A\cos\omega_0 t} d^\dagger d$ and the dissipator
$ \mathcal{D}\bqty{\bullet}=\gamma_l\pqty{d \bullet d^\dagger -\frac{1}{2}\acomm{d^\dagger d}{\bullet}}+\gamma_p\pqty{d^\dagger \bullet d -\frac{1}{2}\acomm{d d^\dagger}{\bullet}}$
we write the vectorized time-dependent Lindbladian in the superfermion representation. Written in the number basis $\ket*{M,\tilde{N}}=\ket*{M}\otimes\ket*{\tilde{N}},$ the Lindbladian is
\begin{widetext}
\begin{equation}
\mathcal{L}(t)=\left(\begin{array}{c|@{}c c@{}}
&  \mqty{\ket{0,\tilde{0}} & \ket{1,\tilde{1}}}  & \mqty{\ket{1,\tilde{0}} & \ket{0,\tilde{1}}} \\
\hline
\mqty{\ket{0,\tilde{0}}\\ \ket{1,\tilde{1}}} & \begin{matrix}
  -\gamma_p &  \gamma_l \\
  \gamma_p & -\gamma_l
  \end{matrix}
  &  \mathbf{0} \\
\mqty{\ket{1,\tilde{0}} \\ \ket{0,\tilde{1}}} &   \mathbf{0} &
  \begin{matrix}
  -i\varepsilon(t)-\gamma/2 &             0           \\
            0         & i\varepsilon(t)-\gamma/2
  \end{matrix}
\end{array}\right)\equiv \mqty(\mathbf{M_0} & \mathbf{0} \\ \mathbf{0} & \mathbf{M_1(t)})
\end{equation}
\end{widetext}
where $\gamma_l=\sum_\alpha \gamma^{-}_{\alpha}$, $\gamma_p=\sum_\alpha \gamma^{+}_{\alpha},$ and $\gamma=\gamma_l+\gamma_p=\sum_\alpha \gamma_\alpha$ are the coefficients given in  Sec.~\ref{sec:RLM}. As explained in the main text, the Lindbladian commutes with $\mathcal{K}=i(N-\tilde{N}),$ the operator that counts the difference between the number of physical and tilde fermions. As a result, it has a block-diagonal structure with each block labeled by the eigenvalues $K$ of the operator $\mathcal{K}.$ We can therefore calculate the evolution superoperator explicitly in the different subspaces. In the $K=0$ sector there is no dependence on time, so $\mathcal{L}^{(0)}(t)=\mathbf{M_0}$ and it follows by simple exponentiation that $V^{(0)}(t,t_0)=e^{(t-t_0)\mathbf{M_0}}.$ In the $\abs{K}=1$ sector, we write $\mathcal{L}^{(1)}(t)=-i(\varepsilon_d+2A\cos\omega_0 t)\sigma_z -\gamma/2$ where $\sigma_z=\mqty(\dmat{1,-1}).$ Exponentiation gives
\begin{widetext}
\begin{equation}
    V^{(1)}(t,t_0)=\mathcal{T}e^{\int_{t_0}^t \dd{s}\mathbf{M_1(s)}}=e^{-(i\varepsilon_d\sigma_z+\gamma/2)(t-t_0)} e^{-i\frac{2A}{\omega_0}(\sin\omega_0 t-\sin\omega_0 t_0)\sigma_z}.
\end{equation}
Using the Jacobi-Anger identity, the exponential of the sine function can be expanded in terms of Bessel functions of the first kind
\begin{equation}
    e^{-i\frac{2A}{\omega_0}\sin\omega_0 t\sigma_z}=\cos(\frac{2A}{\omega_0}\sin\omega_0 t)-i\sin(\frac{2A}{\omega_0}\sin\omega_0 t)\sigma_z=\sum_{n=-\infty}^{+\infty} J_n\pqty{\frac{2A}{\omega_0}} e^{-in\omega_0 t}(\sigma_z)^n,
\end{equation}
so that 
\begin{equation}
    V^{(1)}(t,t_0)=e^{-(i\varepsilon_d\sigma_z+\gamma/2)(t-t_0)}\sum_{m,n} J_n\pqty{\frac{2A}{\omega_0}}J_m\pqty{\frac{2A}{\omega_0}} e^{-in\omega_0 t} e^{im\omega_0 t_0}(\sigma_z)^{m+n}.
\end{equation}
The calculation of the retarded Green's function $ G^r(t,t_0)=-i\theta(t-t_0)\expval{\acomm{d(t)}{d^\dagger(t_0)}}$ follows through the quantum regression theorem
\begin{equation}
    G^r(t,t_0)=-i\theta(t-t_0)\pqty{\mel{\mathds{1}}{d V(t,t_0) d^\dagger}{\rho(t_0)}+\mel{\mathds{1}}{d^\dagger V(t,t_0) d}{\rho(t_0)}^\ast}
\end{equation}
As explained in the main text, the steady state is found in the $K=0$ sector and corresponds to the nullspace of $\mathbf{M_0}$ $\ket{\rho_{ss}}=\ket{\phi_0^0}=\mqty(\gamma_l/\gamma_p& 1 & 0 & 0)^T$ while the vectorized identity is the left zero eigenvector $\bra*{\mathds{1}}=\bra*{\tilde{\phi}_0^0}=\frac{\gamma_p}{\gamma_l+\gamma_p}\mqty(1& 1 & 0 & 0).$ We find the retarded Green's function in the steady state 
\begin{equation}\label{app:eq:Gr}
    G^r(t,t_0)=-i\theta(t-t_0)e^{-(i\varepsilon_d+\gamma/2)(t-t_0)}
    \sum_{m,n} J_n\pqty{\frac{2A}{\omega_0}}J_m\pqty{\frac{2A}{\omega_0}} e^{-in\omega_0 t} e^{im\omega_0 t_0}.
\end{equation}
\end{widetext}
In Wigner representation, we obtain
\begin{equation}
    G^r_l(\omega)=\sum_m \frac{J_m\pqty{\frac{2A}{\omega_0}}J_{m+l}\pqty{\frac{2A}{\omega_0}}}{\omega-\varepsilon_d-(m+\frac{l}{2})\omega_0+i\frac{\gamma}{2}}
\end{equation}
and the time-averaged Green's function at $l=0$
\begin{equation}
    G^r_0(\omega)=\sum_m \frac{J^2_m\pqty{\frac{2A}{\omega_0}}}{\omega-\varepsilon_d-m\omega_0+i\frac{\gamma}{2}}.
\end{equation}
In the wide-band limit, and assuming the couplings to the leads are proportional, the time-averaged current from the left lead to the resonant level in the steady state is~\cite{jauho_timedependent_1994,PLATERO20041}
\begin{equation}
    \expval{I_L}=-2\!\int\frac{\dd{\omega}}{2\pi}\bqty{f_L(\omega)\!-\!f_R(\omega)}\frac{\gamma_L\gamma_R}{\gamma_L\!+\!\gamma_R}\Im\expval{A(\omega,t)}
\end{equation}

where the function $A(\omega,t)\equiv\int\dd{t_0}G^r(t,t_0) e^{i\omega(t-t_0)}$ can be calculated by straightforward integration of Eq.~(\ref{app:eq:Gr}). We find

\begin{equation}
 \expval{A(\omega,t)}=\sum_{m}\frac{J^2_m\pqty{\frac{2A}{\omega_0}}}{\omega-\varepsilon_d-m\omega_0+i\frac{\gamma}{2}}.
\end{equation}

Finally, the current is given by 

\begin{equation}
\begin{split}
    \expval{I_L}&=\int\frac{\dd{\omega}}{2\pi}\bqty{f_L(\omega)-f_R(\omega)}\\
    &\times \sum_{m}J^2_m\pqty{\frac{2A}{\omega_0}}\frac{\gamma_L\gamma_R}{(\omega\!-\!\varepsilon_d\!-\!m\omega_0)^2+\frac{\gamma^2}{4}}  .  
\end{split}
\end{equation}

\bibliography{biblio}

\end{document}